\documentclass[twocolumn,superscriptaddress,floatfix,preprintnumbers, nofootinbib,hyperref]{revtex4} 
\pdfoutput=1
\usepackage[colorlinks=true,breaklinks=true]{hyperref}
\usepackage[normalem]{ulem}
\usepackage[utf8]{inputenc}
\hypersetup{allcolors=[rgb]{0.0 0.0 0.6},linkcolor=[rgb]{0.75 0.05 0.05}}
\usepackage{amsmath,amssymb}
\usepackage{epsfig}  
\usepackage{graphicx}  
\usepackage{subfigure}
\usepackage{slashed}       
\usepackage{url}
\usepackage{color}
\usepackage{multirow}
\usepackage{comment}

\hypersetup{allcolors=[rgb]{0.0 0.0 0.6},linkcolor=[rgb]{0.75 0.05 0.05}}
\usepackage[dvipsnames]{xcolor}

\DeclareMathOperator{\GeV}{GeV}

\DeclareMathOperator{\MeV}{MeV}

\DeclareMathOperator{\s}{s}

\DeclareMathOperator{\erg}{erg}

\DeclareMathOperator{\kpc}{kpc}
\DeclareMathOperator{\g}{g}

\DeclareMathOperator{\kton}{kton}

\newcommand{\bk}{{\bf k}}

\newcommand{\bp}{{\bf p}}
\newcommand{\bP}{{\bf P}}

\newcommand{\beq}{\begin{equation}}
\newcommand{\eeq}{\end{equation}}

\allowdisplaybreaks

\bibpunct{[}{]}{,}{n}{}{,}
\definecolor{ForestGreen}{RGB}{34,139,34}

\newcommand{\X}{{\cal X}}

\begin{document}

\title{
Axion signatures from supernova explosions \\
through the nucleon electric-dipole portal}

\author{Giuseppe Lucente}
\email{giuseppe.lucente@ba.infn.it}
\affiliation{Dipartimento Interateneo di Fisica “Michelangelo Merlin”, Via Amendola 173, 70126 Bari, Italy}
\affiliation{Istituto Nazionale di Fisica Nucleare - Sezione di Bari, Via Orabona 4, 70126 Bari, Italy}%

\author{Leonardo Mastrototaro}
\email{lmastrototaro@unisa.it}
\affiliation{{Dipartimento di Fisica ``E.R. Caianiello'', Università degli Studi di Salerno, Via Giovanni Paolo II, 132 - 84084 Fisciano (SA), Italy}}
\affiliation{Istituto Nazionale di Fisica Nucleare, Sezione di Napoli, Gruppo Collegato di Salerno, Via Giovanni Paolo II, 132 I-84084 Fisciano, Salerno, Italy.}

\author{Pierluca Carenza}\email{pierluca.carenza@fysik.su.se}
\affiliation{The Oskar Klein Centre, Department of Physics, Stockholm University, Stockholm 106 91, Sweden
}

\author{Luca Di Luzio}
\email{luca.diluzio@unipd.it}
\affiliation{Dipartimento di Fisica e Astronomia `G.~Galilei', Universit\`a di Padova, Via Marzolo 8, I-35131 Padova, Italy}
\affiliation{Istituto Nazionale Fisica Nucleare, Sezione di Padova, Via Marzolo 8, I-35131 Padova, Italy}

\author{Maurizio Giannotti}\email{mgiannotti@barry.edu}
\affiliation{Department of Chemistry and Physics, Barry University, 11300 NE 2nd Ave., Miami Shores, FL 33161, USA}%

\author{Alessandro Mirizzi}
\email{alessandro.mirizzi@ba.infn.it}
\affiliation{Dipartimento Interateneo di Fisica “Michelangelo Merlin”, Via Amendola 173, 70126 Bari, Italy}
\affiliation{Istituto Nazionale di Fisica Nucleare - Sezione di Bari, Via Orabona 4, 70126 Bari, Italy}%

\date{\today}
\smallskip

\begin{abstract}
We consider axions coupled to nucleons and photons only through the nucleon electric-dipole moment (EDM) portal. This coupling is a model-independent feature of QCD axions, which solve the strong CP problem, and might arise as well in more general axion-like particle setups. We revise the supernova (SN) axion emission induced by the nucleon EDM coupling and refine accordingly the SN 1987A bound. Furthermore, we calculate the axion flux from a future Galactic SN and show that it might produce a peculiar and potentially detectable gamma-ray signal in a large underground neutrino detector such as the proposed Hyper-Kamiokande. The possibility to detect such a signal offers a way to search for an oscillating nucleon EDM complementary to CASPERe, without relying on the assumption that axions are a sizeable component of the dark matter. Furthermore, if axions from SN produce an observable signal, they could also lead to an amount of cosmological extra radiation observable in future cosmic surveys.
\end{abstract}

\maketitle

\section{Introduction}

Axions are (pseudo)-scalar fields 
predicted in many well-motivated extensions of the Standard Model (SM)~\cite{Jaeckel:2010ni,DiLuzio:2020wdo}. The most notable example is the QCD axion~\cite{Weinberg:1977ma,Wilczek:1977pj}, which emerges as an essential ingredient in the Peccei-Quinn (PQ) solution of the strong CP problem~\cite{Peccei:1977hh,Peccei:1977ur}. More generally, in the context of quantum field theory, axions emerge naturally as the Goldstone bosons of global symmetries that are broken at some high scale $f_a$~\cite{Peccei:1977hh,Weinberg:1977ma,Wilczek:1977pj,Masso:2004cv}. 
Ultra-light axions also appear in other frameworks such as supergravity or string theory~\cite{Arvanitaki:2009fg,Cicoli:2012sz,Broeckel:2021dpz}. Besides theoretical motivations, there is a huge attention towards axions since these are excellent candidates to account for some or all of the dark matter that we observe in the Universe~\cite{Preskill:1982cy,Abbott:1982af,Dine:1982ah}. 

Low-energy experimental tests depend on the axion effective couplings to photons and matter fields, notably 
\beq
\label{eq:interactions}
\mathcal{L}_a = C_{a\gamma} \frac{\alpha}{8\pi} \frac{a}{f_a} F_{\mu\nu} \tilde F^{\mu\nu} + C_{a\Psi} \frac{\partial_\mu a}{2 f_a} \bar \Psi \gamma^\mu \gamma_5 \Psi + \ldots \, ,
\eeq
where $F_{\mu\nu}$ ($\tilde F^{\mu\nu}$) denotes the electromagnetic field strength (and its dual), $\psi = e,p,n$ runs over low-energy matter fields, and $C_{a\gamma,\Psi}$ are naturally expected to be  $\mathcal{O}(1)$ adimensional coefficients. 

From a phenomenological perspective, it is often assumed the presence of only one of the previous couplings and studied the possibility to constrain each of them separately. The axion-photon coupling (first one in Eq.~\eqref{eq:interactions}) is arguably the most used in experimental searches and phenomenological studies. 
Notably, in the presence of an external magnetic field, the axion-photon interaction leads to the phenomenon of axion-photon mixing~\cite{Raffelt:1987im}. This effect is exploited by several ongoing and upcoming axion search experiments (see~\cite{Irastorza:2018dyq,DiLuzio:2020wdo,Sikivie:2020zpn} for recent reviews). 
The axion-photon coupling would also cause axions to be produced in stellar plasmas via the Primakoff process~\cite{Raffelt:1985nk}. Therefore astrophysical observations of the Sun, globular cluster systems and supernovae (SNe) offer unique sensitivity to axion interactions (see~\cite{Raffelt:2006cw,DiLuzio:2021ysg} for reviews). The axion-fermion couplings in Eq.~(\ref{eq:interactions}) also lead to axion production in different stellar systems, e.g.~via electron  bremsstrahlung in white dwarfs and red giants~\cite{Carenza:2021osu}, or nucleon bremsstrahlung~\cite{Burrows:1988ah,Burrows:1990pk,Carenza:2019pxu} and pion conversion~\cite{Carenza:2021ebx,Fischer:2021jfm} in SNe. Furthermore, experimental techniques sensitive to the fermion couplings have been recently conceived (see e.g.~\cite{Arvanitaki:2014dfa,Barbieri:2016vwg,Crescini:2017uxs,JacksonKimball:2017elr}). 
 
Above the scale of QCD confinement, the axion interactions in Eq.~(\ref{eq:interactions}) stem from an axion effective Lagrangian involving quarks and gluons (as well as other SM fields)  
\beq 
\label{eq:LaaboveQCD}
\mathcal{L}_a = \frac{\alpha_s}{8\pi} \frac{a}{f_a} 
G^a_{\mu\nu} \tilde G^{a\mu\nu} + \ldots \,
\eeq
where $G^a_{\mu\nu}$ ($\tilde G^{a\mu\nu}$) denotes the gluon field strength (and its dual). The axion coupling with gluons is the most generic feature in the case of the QCD axion, introduced in order to solve the strong CP problem. The two-gluon coupling would allow for the gluonic Primakoff effect in analogy to what is expected for the photon coupling. This effect might be relevant for thermal axion production in the primordial hot quark-gluon plasma~\cite{Masso:2002np,Graf:2010tv,Salvio:2013iaa,Baumann:2016wac,DEramo:2021psx,DEramo:2021lgb,Giare:2021cqr,Altherr:1990tf}. At the same time, the axion-gluon vertex is responsible for an irreducible contribution to the axion couplings to photons and nucleons in Eq.~(\ref{eq:interactions}). These couplings, however, receive other equally important contributions dependent on the specific UV completion of the model. It is, thus, possible to conceive QCD axion models in which they are suppressed compared to their natural $\mathcal{O}(1)$ values, as we argue in Appendix \ref{sec:stealthaxions}. In the following, we will assume that the couplings to photons and fermions are suppressed.

Finally, the gluonic vertex induces a model independent \emph{nucleon EDM portal} interaction\footnote{We observe that the axion-gluon coupling in Eq.~(\ref{eq:LaaboveQCD}) is not the only possible ``microscopic'' source for the nucleon EDM portal. In fact, the latter could arise as well from axion-like particle interactions with the basis of CP-violating SM quark and gluon effective operators, such as for example an interaction of the type $a \, \bar q i \gamma_5 \sigma_{\mu\nu} q F^{\mu\nu}$, with no relation to the solution of the strong CP problem. Hence, the axion-nucleon EDM portal in Eq.~(\ref{eq:dipole portal}) describes a more general class of axion-like particle theories and could be decorrelated from the constraints stemming from the axion-gluon operator.} 
 \begin{equation}
    \mathcal{L}_a^{\rm nEDM} =-\frac{i}{2} g_{d,N} a {\bar N} \gamma_5 \sigma_{\mu\nu} N {F}^{\mu\nu} \, ,
    \label{eq:dipole portal}
 \end{equation}
(with $N = p,n$) which leads to an axion-dependent nucleon electric dipole moment (EDM), $d_N=g_{d,N} a$~\cite{Graham:2013gfa}. It should be noted that this interaction is in one-to-one correspondence with the axion-gluon coupling via the relation 
\begin{equation}
g_{d,N} = \frac{C_{aN\gamma}}{m_N f_a} \,\ , 
\label{eq:gd_fa}
\end{equation}
 with $C_{an\gamma} = - C_{ap\gamma} = 0.0033(15)$~\cite{Pospelov:1999ha} (i.e., $g_{d,n}=-g_{d,p}\equiv g_{d}$), and hence it is a \emph{model-independent} feature of QCD axions. The nucleon-EDM interaction is particularly important for axion dark matter searches. Indeed, the oscillating axion dark matter field would imprint the same oscillations into the EDM of protons and neutrons~\cite{Graham:2013gfa}. The detection of such a feature is the ambitious goal of the Cosmic Axion Spin Precession ExpeRiment (CASPERe) experiment~\cite{Budker:2013hfa,JacksonKimball:2017elr}.
Oscillating electric-dipole moments of atoms and 
molecules can also be 
generated by the interaction in Eq.~(\ref{eq:LaaboveQCD})~\cite{Stadnik:2013raa,Flambaum:2019ejc,Roussy:2020ily}.

As we shall see in detail in the discussion below, a strong indirect constraint on the axion interaction with the nucleon-EDM can be derived from the analysis of the SN 1987A observed neutrino signal. This possibility was originally presented in Ref.~\cite{Graham:2013gfa}, which provided a back-of-the-envelope (but, as it turns out, rather accurate) estimate of the axion emission rate from a SN through the $N+\gamma \to N+a$ process. In absence of a direct axion-nucleon coupling, this rate is proportional to $g_d^2$. It is well known that an overly efficient axion rate would reduce the duration of the observed SN 1987A neutrino signal~\cite{Kamiokande-II:1987idp,Bionta:1987qt}, thus allowing to constraint the efficiency of the above process and, consequently, to set a bound on $g_d$.  

Given the relevance of the SN bound to constrain the nucleon-EDM portal, we devote our paper to investigate the bounds and signatures of such interactions from SN observations. The plan of our work is as follows. In Sec.~\ref{sec:Supemiss} we present the calculation of the axion emissivity in a SN via the nucleon-EDM portal. 
In Sec.~\ref{sec:SNbound}, we characterize the bound from SN 1987A.  
In Sec.~\ref{sec:axionHK}, we calculate the axion signal from a Galactic SN in a large underground neutrino detector, like Hyper-Kamiokande~\cite{Abe:2011ts}. 
For completeness, in Sec.~\ref{sec:cosmo} we present an estimate of the cosmological bound on axion thermalization through the $N+\gamma \to N+a$ process. 
Finally, in Sec.~\ref{sec:concl}, we discuss the complementary of our findings with other observables related to the nucleon-EDM portal and we conclude. 
There follow three Appendices. 
In Appendix~\ref{sec:stealthaxions}, we discuss model-building aspects of QCD axions with suppressed couplings to photons and matter fields. 
In Appendix~\ref{sec:appem}, we provide further details on the calculation of the SN axion emissivity while in Appendix~\ref{sec:appapprox} we discuss the non-degenerate nucleon limit.

\section{Supernova axion emissivity via nucleon dipole portal}
\label{sec:Supemiss}

\begin{figure}[t!]
		\vspace{0.cm}
		\includegraphics[width=0.9\columnwidth]{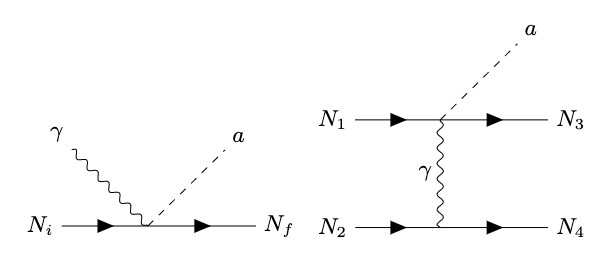}
\caption{Feynman diagrams of the axion production processes: Compton scattering (left) and Bremsstrahlung (right).}
\label{fig:feynm}
	\end{figure}

Axions can be produced in SN through the nucleon dipole portal of Eq.~(\ref{eq:dipole portal}). The two processes which contribute to the rate are the Compton scattering~\cite{Graham:2013gfa}, $N+\gamma \to N+a$, and the nucleon bremsstrahlung process, $N+N \to N+N+a$.\footnote{Since in this model the axion-nucleon coupling is suppressed, this process is different from the nucleon-nucleon bremsstrahlung~\cite{Burrows:1988ah,Burrows:1990pk,Carenza:2020cis}, which is not considered here.} Both processes are shown in Fig.~\ref{fig:feynm}. Of these, the Compton scattering gives the largest contribution while the bremsstrahlung is suppressed by more than one order of magnitude, as further discussed in Appendix~\ref{sec:appapprox}. Therefore, in this section we present only a discussion of the Compton effect. Nevertheless, since the bremsstrahlung contribution has never been considered before in the literature, we provide a detailed derivation of the axion emission rate associated with it in Appendix~\ref{sec:appapprox}.

The matrix element for the Compton process involving a nucleon with a real photon in the initial state (see the left panel in Fig.~\ref{fig:feynm}) is
\begin{equation}
    \mathcal{M}_C=\frac{1}{4}\,g_{d} \bar{u}(p_{f})\,(\gamma^\mu \gamma^\nu-\gamma^\nu \gamma^\mu) \gamma^5 u(p_{i})\, F_{\mu\nu}\;, 
    \label{eq:matrix}
\end{equation}
where $p_f=(E_f, {\bf p}_f)$, $p_i=(E_i, {\bf p}_i)$ are the final and initial nucleon 4-momenta respectively and $g_d\equiv g_{d,n}=-g_{d,p}$ is the coupling to nucleon EDM, with the same magnitude but opposite sign for neutrons $n$ and protons $p$.

In a SN, the photon acquires an effective mass $m_\gamma\approx 16.3~\MeV Y_e^{1/3} \rho_{14}^{1/3}$~\cite{Kopf:1997mv}, where $\rho_{14}= \rho/(10^{14}$ g/cm$^{3})$ and $Y_e$ is the electron fraction. Therefore, we evaluate the spin-averaged squared matrix element from Eq.~\eqref{eq:matrix} assuming the photon as a massive boson (with three degrees of freedom), getting
\begin{align}
    |\overline{\mathcal{M}}|^2&=\frac{g_d^2}{12}|\mathcal{M}|^2 = \nonumber \\
   & =g_d^2 \left[\frac{4}{3} (k\cdot p_f)\,(k\cdot p_i) - \frac{1}{3} m_\gamma^2 (p_f\cdot p_i) + m_N^2 m_\gamma^2\right] \,,
 \label{eq:sqmSN}
\end{align}
where $k$ is the photon 4-momentum and $m_N$ is the nucleon mass.

	\begin{figure*}[t!]
		\vspace{0.cm}
		\includegraphics[width=1.9\columnwidth]{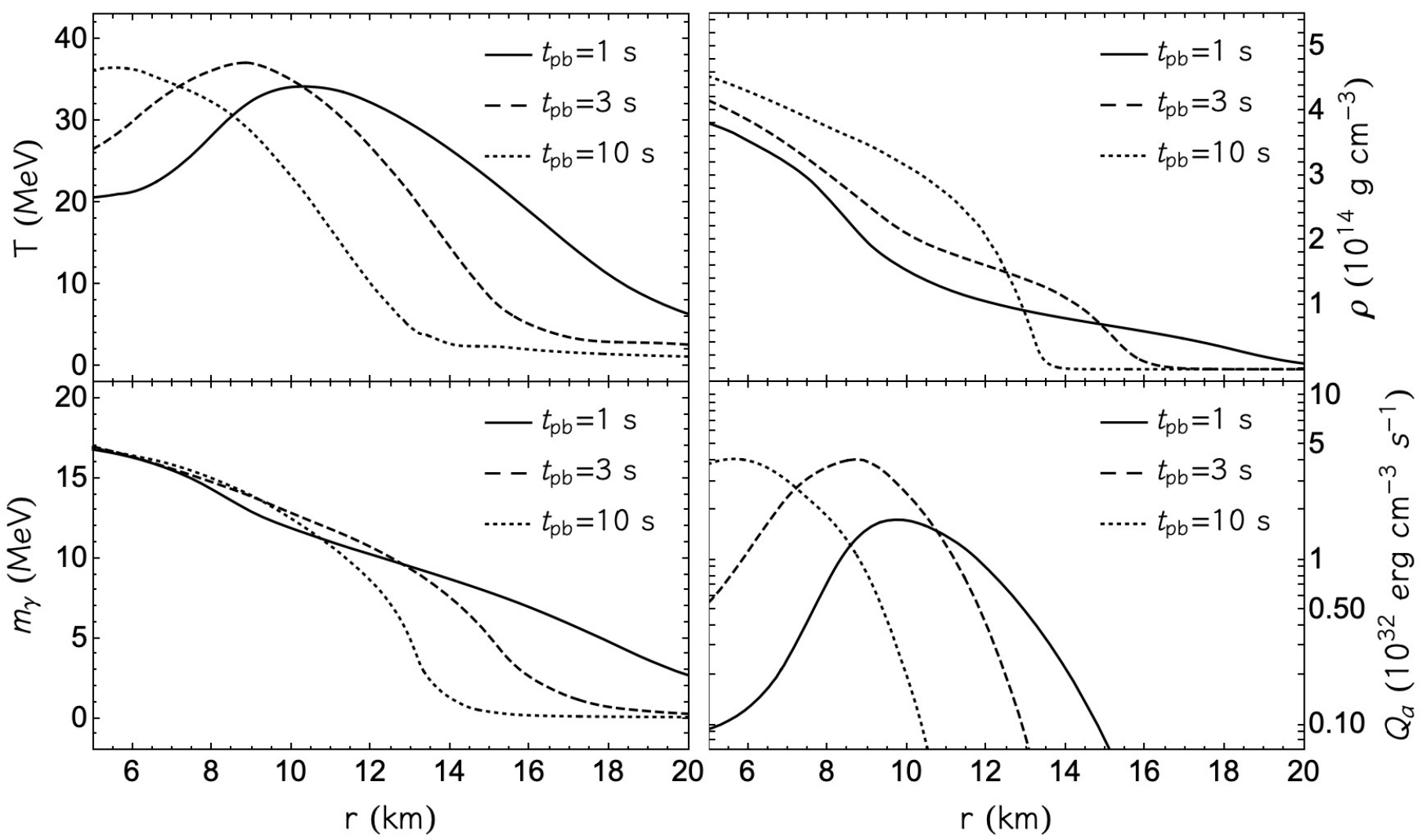}
		\caption{The SN temperature $T$ (upper left panel), the density $\rho$ (upper right panel), the effective photon mass (lower left panel) and the axion emissivity $Q_a$ (lower right panel) as a function of the radius for different post-bounce times $t_{\rm pb}$.
		}
		\label{fig:qafig}
	\end{figure*}
	
The number of axions emitted per unit volume and per unit of time and energy is given by 
\begin{eqnarray}
\frac{d {\dot n_a}}{d E_a} &=& \sum_{\rm nucleons} \int \frac{2 d^3 \bp_i}{(2\pi)^3 2 E_i}\frac{2 d^3 \bp_f}{(2\pi)^3 2 E_f}\frac{3 d^3 \bk}{(2\pi)^3 2 E_k} \frac{ 4 \pi 
E_a^2}{(2\pi)^3 2 E_a} \nonumber \\
 &\times&   (2\pi)^4\delta^{4}(p_{f}+p_{a}-p_{i}-k)|\overline{\mathcal{M}}|^2 f_{p_i} f_{k} (1-f_{p_f})\,, \nonumber \\
 \label{eq:numbdens}
\end{eqnarray}
where $|\overline{\mathcal{M}}|^2$ is given by Eq.~\eqref{eq:sqmSN}, and the distribution functions of the different interacting species are the usual Fermi-Dirac or Bose-Einstein distribution,
\begin{align}
f_i(E)=\frac{1}{e^{\left[ E_i(p_i)-\mu_i\right] /T} \pm1}\;,
\end{align}
where the $+$ sign applies to fermions, the $-$  is for bosons, and $\mu_i$ are the chemical potentials for $i=p,n$, while photons have vanishing chemical potential. Corrections to the dispersion relations $E_i(p_i)$ of nucleons are incorporated through the equation~\cite{Hempel:2014ssa,Martinez-Pinedo:2012eaj}
\beq
E_i = m_N + \frac{{|{\bf p}_i|}^{2}}{2 m^{\ast}_N} +U_i \,\ ,
\label{eq:energy}
\eeq
where $U_i$ is the non-relativistic mean-field potential and $m^{\ast}_N$ is the effective nucleon mass in medium (see Ref.~\cite{Carenza:2019pxu} for details). For definiteness, we take as benchmark for all the different input necessary to characterize the axion emission the  SN  model with $18$ $M_{\odot}$ progenitor simulated in spherical symmetry with the AGILE-BOLTZTRAN code~\cite{Mezzacappa:1993gn,Liebendoerfer:2002xn}.

The differential axion number luminosity, which is defined to be the total number of axions emitted in a specified energy range per unit time from the SN is obtained by integrating Eq.~(\ref{eq:numbdens}) over the SN volume and is given by
\begin{equation}
\frac{d{\cal {\dot N}}_a}{d E_{a}} =\int d^{3}r\,\frac{d{\dot n}_a}{d E_{a}}\;.  
\label{eq:dNA}
\end{equation}
The energy radiated in axions per unit volume and time, called the axion emissivity, can be calculated directly from Eq.~(\ref{eq:numbdens}) as~\cite{Brinkmann:1988vi}
\begin{equation}
Q_{a}=\int d E_{a} E_{a}\frac{d{\dot n}_a}{d E_{a}}\,.
\label{eq:emiss}
\end{equation}
Phase space integration is performed following the procedure of Refs.~\cite{Hannestad:1995rs,Mastrototaro:2021wzl}, as documented in Appendix~\ref{sec:appem}.

In Fig.~\ref{fig:qafig} we show the axion emissivity as a function of the SN radius $r$ at different post-bounce times $t_{\rm pb}$ (bottom right panel), together with the physical properties which determine it, namely the temperature $T$ (upper left panel), the matter density $\rho\sim O(10^{14})$~g~cm$^{-3}$ (upper right panel) and the effective photon mass $m_\gamma \sim 15$~MeV in the core (bottom left panel). In particular, at $t_{\rm pb}=1$~s, the production zone is at $r\sim 10$~km and it moves towards the star center at larger times, reflecting the behaviour of the peak temperature $T_{\rm max} \sim 30-40$~MeV and showing the strong temperature dependence of the production rate. The axion energy luminosity, i.e.~the energy emitted by axions per unit time, is obtained by integrating the emissivity over the stellar volume, i.e.
\begin{equation}
L_{a}= 4\pi \int dr\, r^2\, Q_{a}(r)\,.
\label{eq:ladef}
\end{equation}
We mention that redshift corrections need to be considered in order to evaluate the luminosity for a distant observer, as discussed in Refs.~\cite{Caputo:2021rux,Caputo:2022mah}. Indeed, after its emission, an axion will suffer a gravitational redshift before reaching an observer at infinity. This effect is encoded in the ``lapse'' factor $\alpha$ listed at each radius in the SN simulation data~\cite{Liebendoerfer:2002xn}. This means that the observed axion energy at infinity is $E_{\rm obs} = E_{\rm loc} \times \alpha$, where $E_{\rm loc}$ is the axion energy in the local comoving frame of reference, in which SN-simulation data are provided. In addition, for the rate of emission another redshift correction is required, since the proper time lapse of a comoving observer is related to the distant observer time by the lapse function $\alpha$~\cite{Liebendoerfer:2002xn}. Therefore, the contribution from local emission to the luminosity at infinity can be evaluated by including a factor $\alpha^2$. Moreover, since all physical properties of the star are given in the comoving reference frame of the emitting medium, a Doppler shift effect $\propto (1+2 v_r)$ has to be considered, where $v_r$ is the radial velocity of the medium, which is always very small $|v_r|\ll 1$~\cite{Caputo:2021rux,Caputo:2022mah}. For this reason, the observed axion luminosity at infinity is given by
\begin{equation}
    L_{\rm obs} = 4 \pi \int dr\, r^2 Q_{a} (r)\,\alpha^2 (r) (1+2\,v_r)\,,
\end{equation}
where $Q_a$ is the emission rate evaluated in the local comoving frame of reference. We stress again that since $v_r\ll 1$, the last term in brackets has a small impact on the observed luminosity, while the $\alpha$ factor reduces the luminosity by a factor $\sim 20-30\%$, in agreement with Ref.~\cite{Caputo:2021rux}. 

\section{SN 1987A cooling bound}
\label{sec:SNbound}

The observation of the SN 1987A neutrino burst permits to constrain all the exotic energy losses that would significantly shorten its duration. For quantitative estimates, it is normally assumed that the luminosity associated with exotic processes, calculated at a representative time $t_{\rm pb}=1$ s after the core-bounce, does not exceed the neutrino luminosity $L_\nu \simeq 3\times 10^{52}$~erg~s$^{-1}$~\cite{Raffelt:1987yt,Caputo:2021rux}. In order to place a bound on the axion coupling $g_d$, we evaluate the axion luminosity  adopting  the ``modified luminosity criterium'', (see~\cite{Chang:2016ntp,Lucente:2020whw,Caputo:2021rux})
\begin{equation}
L_{a}= 4\pi \int_0^{R_{\rm p}} dr\, r^2\, \alpha^2 \int dE_a E_a \frac{d{\dot n}_a}{d E_{a}} \langle e^{-\tau(E_a',r)}\rangle\,,
\label{eq:la}
\end{equation}
where the integral of the axion emissivity is performed on the emission region with $R_{\rm p}=40$~km and the exponential suppression $e^{-\tau}$ takes into account the possibility of axion absorption. In particular, $\langle\,e^{-\tau}\rangle$ is a directional average of the absorption factor
\begin{equation}
    \langle e^{-\tau(E_a',r)} \rangle = \frac{1}{2} \int_{-1}^{+1} d\mu\, e^{-\int_0^{\infty} ds \lambda^{-1} (E_a',\sqrt{r^2+s^2+2\,r\,s\,\mu})}\,,
\end{equation}
where $\lambda$ is the axion mean-free path calculated in Eq.~\eqref{eq:lambda} in Appendix~\ref{sec:appem}, $E_a'=E_a \,\alpha(r)/\alpha\left(\sqrt{r^2+s^2+2\,r\,s\,\mu}\right)$ is the axion redshifted energy, $\mu=\cos\beta$ and $\beta$ is the angle between the outward
radial direction and a given ray of propagation along
which $ds$ is integrated.
 
\begin{figure}[t!]
		\vspace{0.cm}
		\includegraphics[width=0.95\columnwidth]{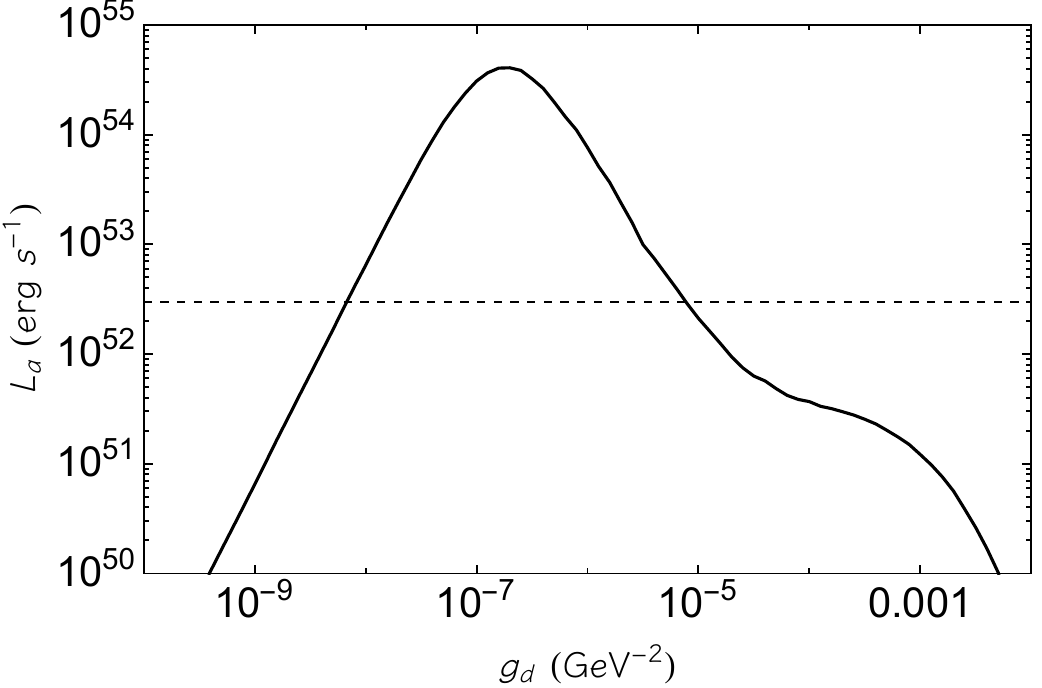}
		\caption{Dependence of $L_a$ on the coupling strength $g_d$ at $t_{pb}=1$~s.
		The horizontal dashed line denotes the neutrino luminosity $L_\nu=3\times 10^{52}$~erg~s$^{-1}$. Couplings giving $L_a \gtrsim L_\nu$ are excluded.}
		\label{fig:bound}
	\end{figure}

In Fig.~\ref{fig:bound}, we show the expected bound on $g_{d}$ in the $L_a$ \emph{vs} $g_d$ plane. The trend is a typical one often discussed in literature (see, e.g., Ref.~\cite{Raffelt:1987yt}). The region for $g_d \lesssim 10^{-7}\GeV^{-2}$ corresponds to the free-streaming case, where the axion production is dominated by a volume emission and $L_a \propto g_d^2$. Conversely, for $g_d \gtrsim 10^{-6}\GeV^{-2}$ axions enter the trapping regime, where the luminosity is dominated by a surface black-body emission from an ``axion-sphere'' with radius $r_a$, where $L_a \propto r_a^2 T(r_a)^4$ that is a rapidly decreasing function of $r$ so that $L_a$ decreases when $g_d$ increases. We exclude values of $g_{d}$ for which $L_a\gtrsim 3\times 10^{52}$~erg~s$^{-1}$, corresponding to the range $6.7\times 10^{-9}\GeV^{-2}\lesssim g_{d} \lesssim  7.7\times 10^{-6}\GeV^{-2}$. We notice that the bound on $g_d$ in the free-streaming regime is slightly weaker than the simple back-of-the-envelope estimation presented in Ref.~\cite{Graham:2013gfa}, namely $g_{d}\lesssim 4\times 10^{-9}$~GeV$^{-2}$. Furthermore, as shown in Sec.~\ref{sec:cosmo}, values of $g_d$ larger than what excluded by the trapping limit are excluded by the extra radiation produced by the thermalization of axions in the early Universe. Therefore, in the next section we will focus on couplings below the free-streaming bound.

\section{Axion signal in Hyper-Kamiokande}
\label{sec:axionHK}

\begin{figure}[t!]
		\vspace{0.1cm}
		\includegraphics[width=0.95\columnwidth]{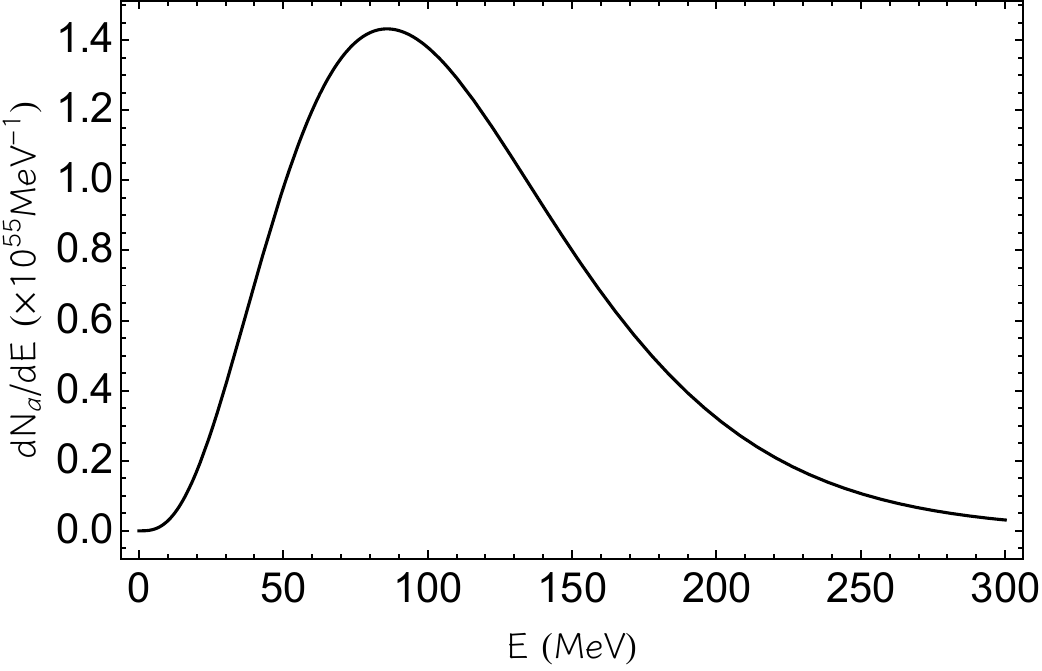}
		\caption{SN axion energy spectrum for $g_{d}=6\times 10^{-9}~\GeV^{-2}$ and $m_{a}\ll T$.}
		\label{fig:flux}
	\end{figure}

Having calculated the SN axion spectrum produced through the nucleon dipole portal, our goal in this section is to discuss detection possibilities from a Galactic SN explosion with next generation neutrino detectors (see, e.g., Ref.~\cite{Mirizzi:2015eza} for a review). For definiteness, we focus on the neutrino underground water Cherenkov detector Hyper-Kamiokande, with a proposed fiducial mass of 374~kton~\cite{Abe:2011ts}. In this case the detection channel is the scattering of the SN axions on free protons in the water
\begin{equation}
a+p \to p+\gamma \,\ ,
\label{eq:detproc}
\end{equation}
producing a visible photon flux. In order to calculate the event rate in Hyper-Kamiokande, one has to consider the  SN  axion fluence  from Eq.~(\ref{eq:dNA}), including gravitational redshift. This is well-represented by the following quasi-thermal spectrum (see also~\cite{Payez:2014xsa})
    \begin{equation}
        \frac{dN_{a}}{dE_{a}}=\left(\frac{g_{d}}{6\times 10^{-9}\GeV^{-2}}\right)^{2}C_{0}\left(\frac{E}{E_{0}}\right)^{\beta}e^{-(1+\beta)\frac{E}{E_{0}}} \,\ ,
    \end{equation}
where $C_{0}=7.49\times10^{56}~\MeV^{-1}$, $E_{0}=113.73~\MeV$ and $\beta=3.09$. This spectrum is shown in Fig.~\ref{fig:flux} for $g_d=6\times 10^{-9}$~GeV$^{-1}$ and negligible axion mass.

The detection cross section, associated with the process in Eq.~(\ref{eq:detproc}), is given by 
\begin{equation}
\begin{split}
\sigma_a&=\frac{1}{2E_{a}}\frac{1}{2m_{N}}\int\frac{2d^{3}p_{f}}{2E_f(2\pi)^3}\frac{2 d^3 k}{2\omega(2\pi)^3}\\
    &\times (2\pi)^4 \delta^{4}(p_a+p_i-k-p_f) |\overline{\mathcal{M}}|^2 \\
    &=\frac{1}{4E_{a}m_{N}(2\pi)^{2}}\int d^{4}k\delta(k^{2})\\
    &\times \delta^{4}(p_a+p_i-k-p_f) |\overline{\mathcal{M}}|^2 p_{f}dE_{f}\\
    &=\frac{g_{d}^{2}E_{a}^{2}}{2\pi}\,\ ,
    \end{split}
\label{eq:crosssec}
\end{equation}
where in the last step we used the  small axion mass limit and the non-relativistic approximation for nucleons, so that $E_\gamma=E_{a}$ and $E_{i}=E_{f}=m_{N}$, subject to the kinematical constraint
\begin{equation}
    E_{f}\le \frac{2E_{a}^{2}+2E_{a}m_{N}+m_{N}^{2}}{2E_{a}+m_{N}} \,\ .
\end{equation}
The produced photon energy spectrum is given by
\begin{equation}
    \frac{d N_\gamma}{d E_\gamma} =  \frac{N_{\rm t}}{4\pi d^{2}} \frac{d N_a}{d E_a}\times \sigma_a(E_a) \,\ , 
\label{eq:dnevde}
\end{equation}
where $d$ is the SN distance from Earth, 
and $N_t$ is the number of targets in the detector, 
    \begin{equation}
        N_{\rm t}= 10^{9} \times N_p \times
        N_A \times \left(\frac{M_{\rm det}}{\rm kton}\right)\times \left(\frac{\textrm{g}/\textrm{mol}}{m_{{\rm H}_{2}{\rm O}}} \right) \,\ ,
    \end{equation}
with $N_p=2$ the number of free protons per water molecule, $N_{\rm A}$ the Avogadro number and $m_{{\rm H}_{2}{\rm O}}=18$~g/mol the molar mass of water.
    
	\begin{figure}[t!]
		\vspace{0.cm}
		\includegraphics[width=0.95\columnwidth]{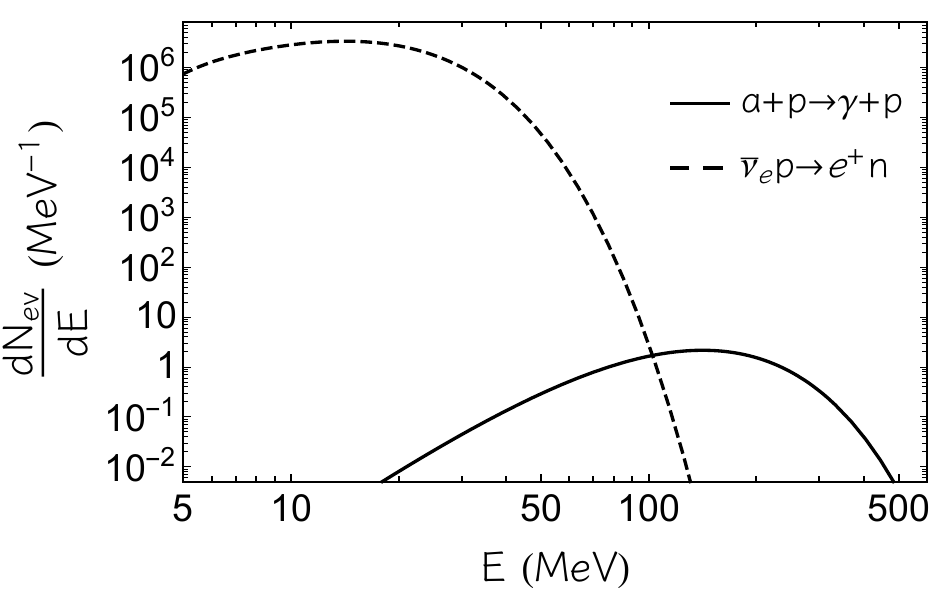}
		\caption{Events rate in Hyper-Kamiokande for the axion signal via $a+p\to p+\gamma$ (continuous curve) for $g_d=6\times10^{-9}$ GeV$^{-2}$ and for $\bar\nu_e$ inverse beta decay $\bar\nu_e+p\to n+e^+$ (dashed curve) for a SN at $d=0.2$~kpc.}
		\label{fig:dNevdE}
	\end{figure}

    	\begin{figure}[t!]
		\vspace{0.1cm}
		\includegraphics[width=0.95\columnwidth]{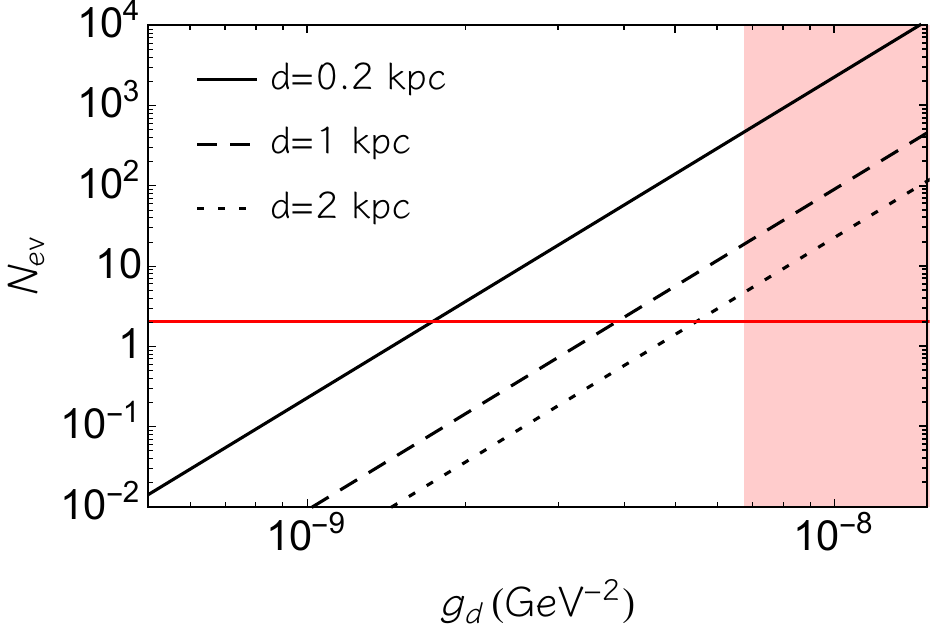}
		\caption{Number of expected axion events with energy $E\gtrsim 100$~MeV in Hyper-Kamiokande as function of $g_d$ for different values of the SN distance. The red line indicates a threshold value of 2 events, required for the detection. The magenta region is excluded by the energy-loss criterion.}
		\label{fig:events}
	\end{figure}
In Fig.~\ref{fig:dNevdE}, we show the event rate in Hyper-Kamiokande for the axion signal via $a+p\to p+\gamma$ (continuous curve) and for a Galactic SN at $d=0.2$~kpc, representative of the distance of the red supergiant star Betelgeuse~\cite{osti_22863070}. 
For comparison, we show the neutrino event rate associated with inverse beta decay process, $\bar\nu_e + p \to n + e^+$ (dashed curve) which is the dominant detection channel for SN neutrinos (see, e.g., Ref.~\cite{Fogli:2004ff}). 
It is interesting to realize that for $E \gtrsim 100$~MeV, the axion signal emerges over the $\bar\nu_e$ background, offering a potential window of detection. 
The high-statistics SN neutrino detection can be used as an external trigger for the axion detection. 
Indeed, it selects a ${\mathcal O}(10)$~s time window to look at the coincidence of at least two photons from axions signal. 
Notably, the accidental background coincidence in a 10~s window is small, less than one event every three years.\footnote{Mark Vagins, private communication.} 
The number of axion events for $E>100$~MeV is given by 
\begin{equation}
 N_{\rm ev}=290 \left(\frac{g_{d}}{6\times 10^{-9}\GeV^{-2}}\right)^{4}\left(\frac{M_{\rm det}}{374\kton}\right)\left(\frac{d}{0.2\kpc}\right)^{-2} \, .
 \end{equation}

    	\begin{figure}[t!]
		\vspace{0.1cm}
		\includegraphics[width=0.95\columnwidth]{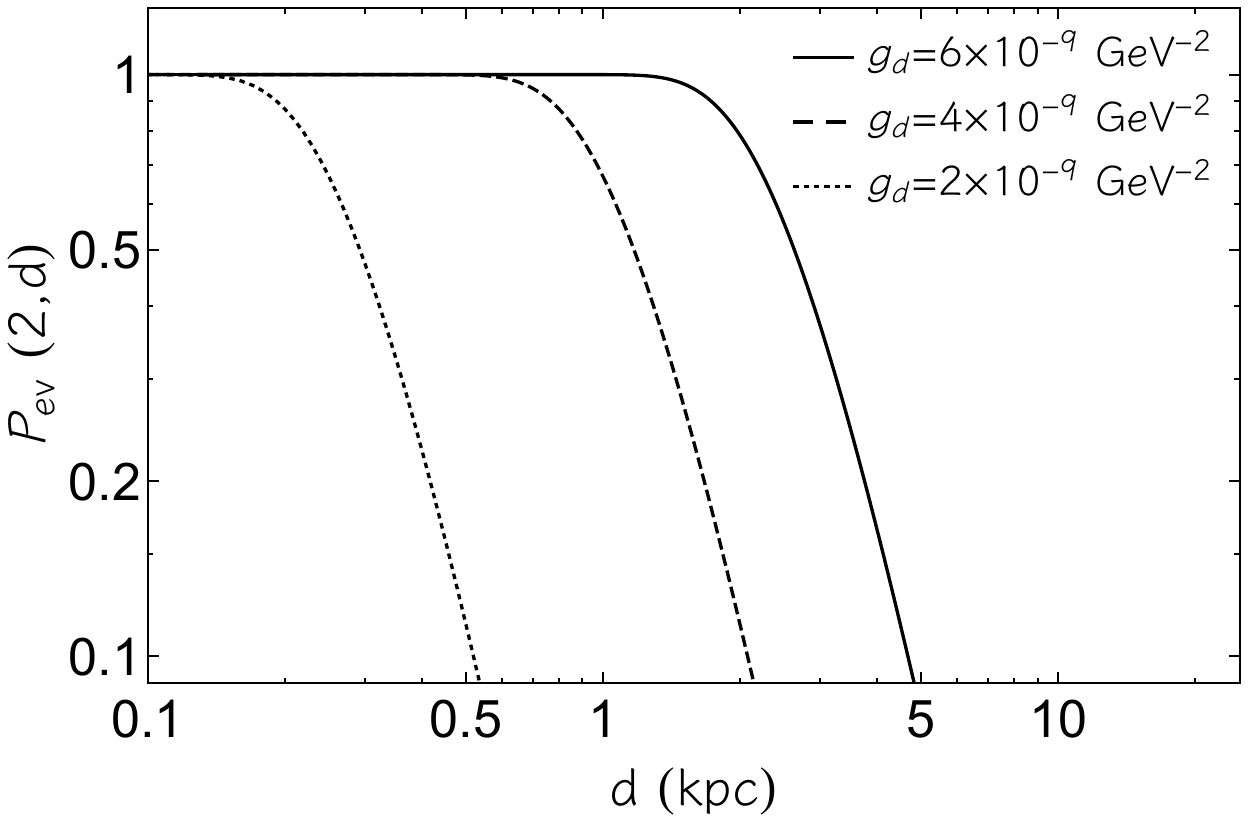}
		\caption{Poisson probability to detect more than 2  axion-induced photon events as a function of the distance, for different values of $g_d$.}
		\label{fig:pev2}
	\end{figure}

In Fig.~\ref{fig:events}, we show the number of expected events in Hyper-Kamiokande as a function of $g_d$, for different values of the SN distance. 
It is apparent that a few hundreds of events would be detected for $g_d\approx 6\times 10^{-9}$~GeV$^{-2}$ near the cooling bound and for a SN explosion at distance $d=0.2$~kpc, such as Betelgeuse. 
Distances up to $d \lesssim 2$~kpc would give a handful of events for the same value of the coupling. 
For a close-by SN at $d \lesssim 0.2$~kpc, we expect to observe few events for couplings larger than $g_d\approx 2\times 10^{-9}$~GeV$^{-2}$. 
In order to quantify the sensitivity to $g_d$ as a function of the SN distance $d$, in Fig.~\ref{fig:pev2} we show the Poisson probability to detect more than two photon events with $E>100$~MeV in Hyper-Kamiokande as a function of the SN distance, for three different values of the coupling, evaluated as
\begin{equation}
    P_{\rm ev} (2, d) = \sum_{n=2}^{\infty} \frac{N_{\rm ev}^n (d)}{n!} e^{-N_{\rm ev}(d)}\,.
\end{equation}
We see that for $g_d = 6 \times 10^{-9}$~GeV$^{-2}$, there is a non-negligible probability ($P_{\rm ev}\gtrsim 0.5$) to detect an axion-signal up to 2.5 kpc. 
For $g_d = 4 \times 10^{-9}$~GeV$^{-2}$, the sensitivity radius is reduced to 1~kpc, and for $g_d = 2 \times 10^{-9}$~GeV$^{-2}$ to 300~pc.
There are $\sim 30$ SN candidates in a radius $d<1$ kpc~\cite{Mukhopadhyay:2020ubs}.
According to our analysis, if one of these goes SN we might expect, together with a huge neutrino signal, a handful of high-energy events associated with the nucleon dipole portal to axions.

\section{Cosmological bounds on axion extra radiation}
{\label{sec:cosmo}}

A complementary constraint on the axion nucleon dipole portal can be derived from measurements of extra radiation in the early universe.
In fact, below the QCD phase transition, the process $N+\gamma \leftrightarrow N+a$  
becomes an effective process to produce a thermal population of axions which would contribute to extra radiation. 
The axion production rate in the early Universe is given by $\Gamma=n_{N}\sigma_{aN\to \gamma N}$ where $n_{N}$ is the nucleon thermal number density and the production cross section $\sigma_{aN\to \gamma N}$  is given by Eq.~(\ref{eq:crosssec}). 
Axions decouple when $\Gamma \simeq H$, where $H$ is the Universe Hubble expansion rate. Having determined the axion decoupling temperature $T_{D}$, it is possible to calculate the effective number of relativistic degrees of freedom, appearing as extra radiation, as~\cite{Kolb:1981hk,DiLuzio:2020wdo}
\beq 
\label{eq:DeltaNeff}
\Delta N_{\rm eff} \simeq 0.027 \left( \frac{106.75}{g_{*,s}(T_{D})} \right)^{4/3} \, ,
\eeq
where $g_{*,s}(T_{D})$ are the entropic effective degrees of freedom (normalized to the total number of SM degrees of freedom). 
The sensitivity of the Planck 2018 data is enough to exclude $\Delta N_{\rm eff}\gtrsim0.35$ at $95\%$ CL~\cite{Planck:2018vyg}, which corresponds to $g_{d}\gtrsim 6\times 10^{-6}\GeV^{-2}$. 
Therefore, the cosmological bound nicely connects with the exclusion given by the SN 1987A in the trapping regime. 
We remark that, for $m_{a}\gtrsim 1$~eV, axions would be too heavy to be considered dark radiation and their constraint from contributing to dark matter is much weaker than the one from $\Delta N_{\rm eff}$.

For values below the SN 1987A bound in the free-streaming regime, $g_{d} \lesssim  7.7\times 10^{-6}\GeV^{-2}$, axions would decouple before the QCD phase transition. In this case the processes relevant for their thermalization are the ones with gluons, rather than with nucleons. In this case the decoupling temperature can be estimated as $T_{D} \simeq 4 \times 10^{11} (f_a / 10^{12} \, \text{GeV})^2$~\cite{DiLuzio:2020wdo} 
(see also \cite{Masso:2002np,Graf:2010tv,Salvio:2013iaa,Baumann:2016wac,DEramo:2021psx,DEramo:2021lgb,Giare:2021cqr}), which is typically well above the electroweak scale. 
Hence, from Eq.~(\ref{eq:DeltaNeff}) it follows that $\Delta N_{\rm eff} \simeq 0.027$, which is in the reach of future CMB-S4 surveys~\cite{Abazajian:2019eic}. Requiring that the temperature of the Universe was high enough to bring the axion into thermal equilibrium, $T_{\rm RH} > T_D$, CMB-S4 data would be able to probe~\cite{Baumann:2016wac}
\beq 
g_d > 1.3 \times 10^{-14} \ \text{GeV}^{-2} \left( 
\frac{T_{\rm RH}}{10^{10} \ \text{GeV}} \right)^{-1/2} \, .
\eeq

\section{Discussion and Conclusions}
\label{sec:concl}

In this work, we provided a careful quantitative investigation of the bounds and signatures of a nucleon dipole portal to axions from a SN explosion.
First, we have revised the axion production channels in a SN. 
The most relevant channels are the Compton and the bremsstrahlung processes, the last of which had never been considered in the previous literature.
We present a detailed calculation of the rates associated with both processes in Appendix~\ref{sec:appapprox}.
We find that the SN 1987A cooling argument provides the limit $g_d \lesssim 6.7\times 10^{-9}\GeV^{-2}$.
Furthermore, we have shown that for values of $g_d$ below this bound and larger than $10^{-9}\GeV^{-2}$ a future Galactic SN explosion within a radius $d \lesssim {\mathcal O} (1)$ kpc would produce a handful of events through the process $a+p\to p+\gamma$ in the Hyper-Kamiokande detector.

In the case of QCD axion, Eq.~\eqref{eq:gd_fa} holds and the bound on $g_{d}$ can be translated into $f_a\gtrsim 5\times 10^{5}$~GeV. However, we stress that in this case the SN bound on $f_a$ due to nucleon-EDM coupling would be weaker than the HB bound due to axion-photon coupling ($f_a \gtrsim 4\times 10^{6}$~GeV)~\cite{Ayala:2014pea} and the SN bound due to axion-nucleon couping ($f_a \gtrsim 4\times 10^8$~GeV)~\cite{Carenza:2019pxu}. Therefore, the nucleon-EDM axion coupling would be the most important one for axion phenomenology only in the case in which both photon- and nucleon- axion couplings are suppressed. As further discussed in Appendix~\ref{sec:stealthaxions}, the required cancellation cannot be achieved with only a single tuning, but further non-trivial assumptions are needed.

    	\begin{figure*}[t!]
    \centering
		\includegraphics[scale=0.61]{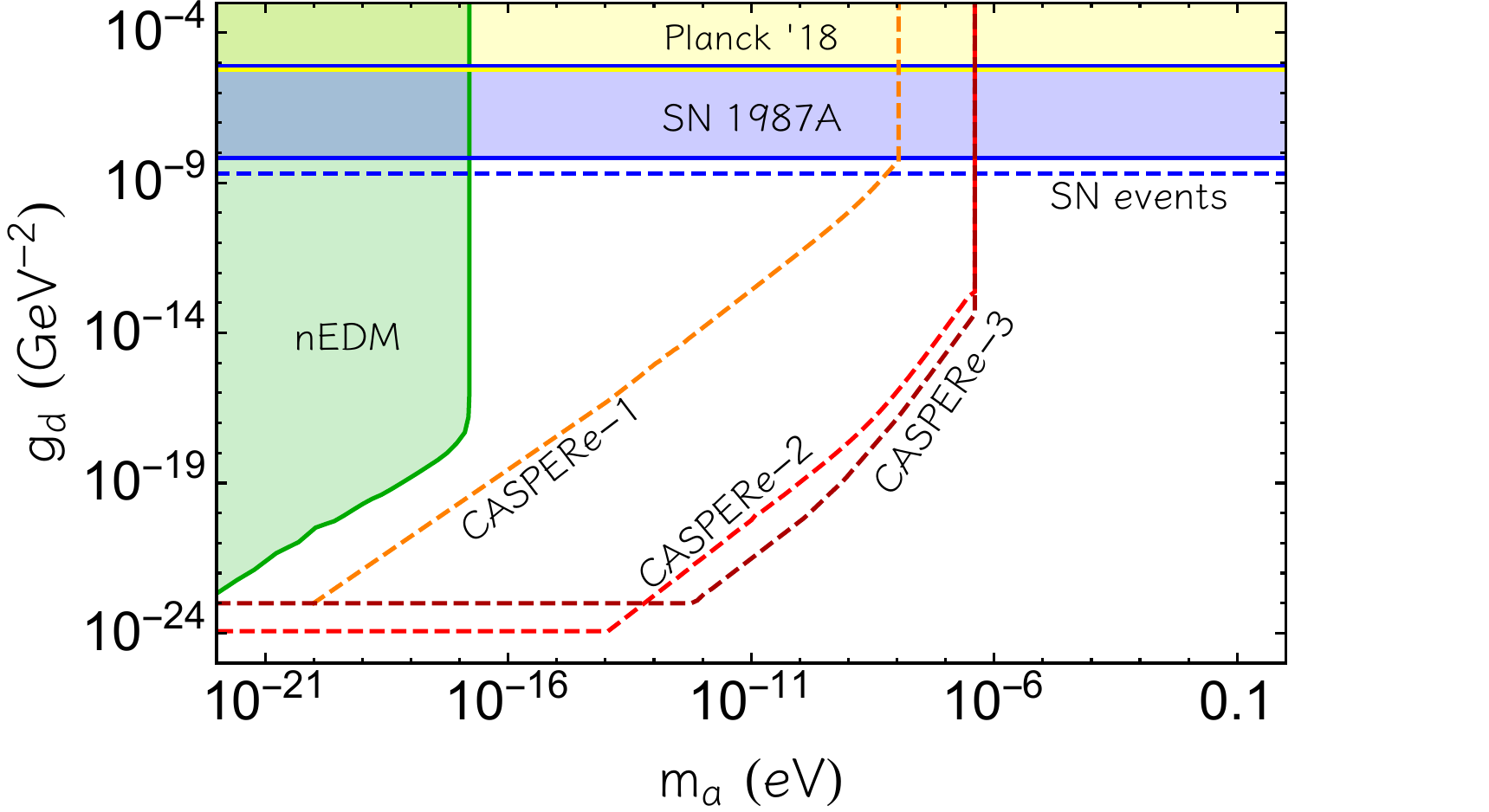} \\
		\vspace{1 cm}
		\includegraphics[scale=0.61]{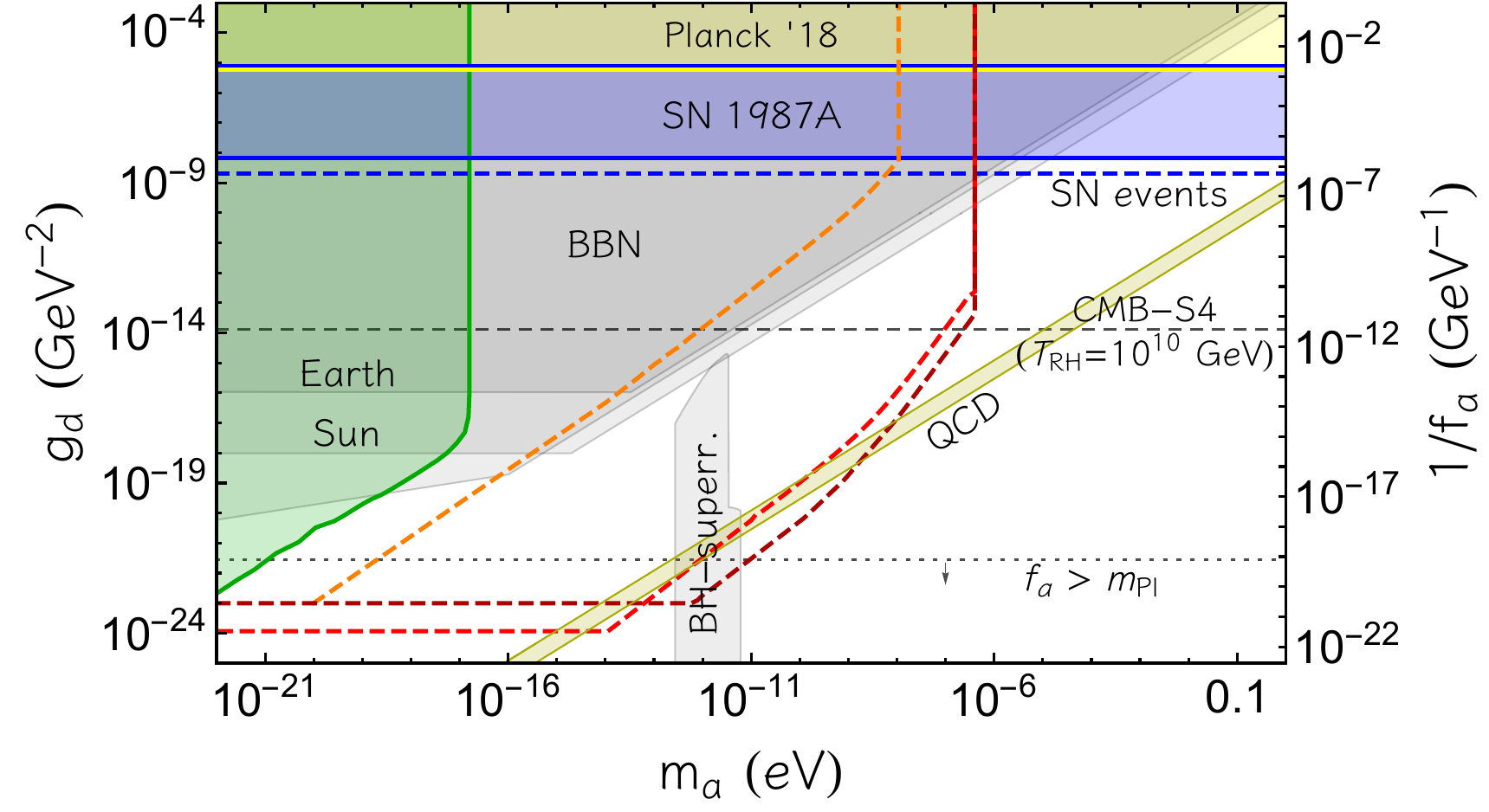}
		\caption{Upper panel: Bounds (full lines) and sensitivity (dashed lines) of future searches in the $g_d$ \emph{vs} $m_a$ plane. Lower panel: Axion parameter space considering interactions derived from the axion-gluon coupling
		[see Eqs.~(\ref{eq:LaaboveQCD})--(\ref{eq:gd_fa})].
		 Bounds (full lines) and future experimental sensitivities (dashed lines) in color pertain to the nucleon-EDM portal, while for the regions in grey we made the further assumption that the origin of the latter coupling is the axion-gluon interaction, with all the other couplings suppressed. 
		 In the region below the dotted line the axion decay constant exceeds the Planck scale. The vertical axis gives the inverse of the decay constant $f_a^{-1}$ on the right and the EDM coupling $g_{d}\propto f_a^{-1}$ on the left.  See the text for a detailed explanation of the plot.
		 \vspace{1 cm}
		 }
		\label{fig:overviewbound}
	\end{figure*}

It is interesting to compare the axion parameter space probed by SNe with sensitivities of other searches in the plane $(g_d,m_a)$, 
as shown in the upper panel of Fig.~\ref{fig:overviewbound}.
There, the shaded green area is excluded by the non-detection of an oscillating nuclear dipole moment in experiments looking for a static one (nEDM, see~\cite{Abel:2017rtm}). The dashed lines represent future sensitivity estimates for different phases of the oscillating EDM experiment CASPERe~\cite{JacksonKimball:2017elr}. Our SN 1987A bound is the blue shaded strip between $6.7\times 10^{-9}\GeV^{-2}\lesssim g_{d} \lesssim  7.7\times 10^{-6}\GeV^{-2}$. Higher values of the coupling are excluded by extra radiation $\Delta N_{\rm eff}$ produced after the QCD phase transition. 
Values $2 \times 10^{-9}\GeV^{-2} \lesssim g_{d} \lesssim  6.7\times 10^{-9}\GeV^{-2}$ (dashed curve) can be probed by axion events from a future close-by Galactic SN explosion. 

In the lower panel 
of Fig.~\ref{fig:overviewbound}, we 
superimpose additional bounds and future sensitivities 
under the assumption that the origin of $g_d$
is 
the anomalous axion-gluon coupling, 
depending on 
$f_a$ in the case of the QCD axion (oblique yellow band)
[see Eqs.~(\ref{eq:LaaboveQCD})--(\ref{eq:gd_fa})]. The region below the dotted grey line ($g_d \lesssim 2.8\times 10^{-22}~\GeV^{-2}$) corresponds 
to $f_a \gtrsim m_{\rm Pl}$, while 
in the grey region (BBN) axions coupled to QCD are inconsistent with the production of the observed abundance of light elements during Big Bang Nucleosynthesis (BBN)~\cite{Blum:2014vsa}. 
Note that both CASPERe and the BBN bound rely on the assumption that the axion comprises the whole cold dark matter. Instead, the bounds denoted as ``Earth'' and ``Sun'' are due to finite density effects. In fact, in models where the axion mass is down-tuned,\footnote{In the presence of an extra dark sector contributing to the axion mass, the latter can be suppressed with respect to the value set by QCD. 
In some studies (e.g.~\cite{Blum:2014vsa,Alonso-Alvarez:2017hsz}) this was supposed to happen via a tuning. 
More recently, exploiting the mechanism proposed in \cite{Hook:2018jle}, Refs.~\cite{DiLuzio:2021pxd,DiLuzio:2021gos} 
showed that the axion mass can be exponentially suppressed in terms of a $Z_N$ symmetry, while solving the strong CP problem and the axion being dark matter. These works motivate the region on the left of the ``QCD axion band'' in Fig.~\ref{fig:overviewbound}.} 
the cancellation of the axion mass can be spoiled in high-density stellar environments where the axion field relaxes to $\langle a \rangle = \pi f_a$, implying various experimental constraints (see Ref.~\cite{Hook:2017psm} for more details). 
The shaded grey region 
around $m_a \sim 10^{-12}$ eV 
represents the most conservative Black-Hole 
superradiance bound \cite{Baryakhtar:2020gao} (but see also Ref.~\cite{Mehta:2020kwu}).
Finally, if axions decouple before the QCD phase transition, e.g.~via the axion-gluon coupling, their contribution to extra radiation would be in the reach of future CMB-S4 observations~\cite{Abazajian:2019eic}, improving over existing constraints on $g_d$. 
In the lower panel of Fig.~\ref{fig:overviewbound} we show what the reach in $g_d$ would be, assuming a reheating temperature of $T_{\rm RH} = 10^{10}$ GeV.

We notice that the region probed by CMB-S4 is complementary to the direct search of axion dark matter by the CASPERe experiment~\cite{JacksonKimball:2017elr}. 
Remarkably, also our SN signal is complementary with CASPERe. Indeed, for masses $m_a \gtrsim 10^{-9}$ eV, if $g_d \gtrsim 10^{-9}\GeV^{-2}$ one would not observe any signal in CASPERe, but there is the possibility to get a few axion-induced  events from a close-by SN and an excess of extra radiation in CMB-S4. 
This is a peculiar scenario where the nucleon dipole portal would be invisible to laboratory experiments and would show up only from the sky.

\begin{acknowledgements}
We warmly thank Andreas Ringwald for interesting discussions stimulating this project. 
AM warmly thanks Mark Vagins for useful discussions on Hyper-Kamiokande during the development of this project.
The work of GL and AM is partially supported by the Italian Istituto Nazionale di Fisica Nucleare (INFN) through the ``Theoretical Astroparticle Physics'' project and by the research grant number 2017W4HA7S ``NAT-NET: Neutrino and Astroparticle Theory Network'' under the program PRIN 2017 funded by the Italian Ministero dell'Universit\`a e della Ricerca (MUR). The work of PC is supported by the European Research Council under Grant No.~742104 and by the Swedish Research Council (VR) under grants  2018-03641 and 2019-02337. The work of LDL is partially supported by the European Union's Horizon 2020 research and innovation programme under the Marie Sk\l odowska-Curie grant agreement No 860881 - HIDDEN. The work of LM is supported by the Italian Istituto Nazionale di Fisica Nucleare (INFN) through the ``QGSKY'' project and by Ministero dell'Istruzione, Universit\`a e Ricerca (MIUR).
\end{acknowledgements}

\appendix

\section{QCD axions with suppressed couplings to photons and matter fields}
\label{sec:stealthaxions}

In this Appendix, we explore the question of whether it is possible to conceive a QCD axion model where the nucleon-EDM portal provides the leading axion interaction. This requires in turn that standard axion couplings to photons and matter fields are suppressed with respect to their natural $\mathcal{O}(1)$ values. 

To formulate the problem in general terms, let us start from the axion effective Lagrangian below the electroweak scale 
\begin{align}
\label{eq:axionEFTEW}
\mathcal{L}_a &= \frac{\alpha_s}{8\pi} \frac{a}{f_a}  G_{\mu\nu}^a \tilde G^{a\mu\nu} 
+ \frac{E}{N} \frac{\alpha}{8\pi} \frac{a}{f_a} F_{\mu\nu} \tilde F^{\mu\nu}
\nonumber \\
&+ \frac{\partial_\mu a}{2 f_a} \bar f c^0_f \gamma^\mu \gamma_5 f + \ldots \, ,
\end{align}
where $E/N$ is the ratio of the QED/QCD anomaly of the PQ current 
and $f = u,d,e,\ldots$ denotes SM Dirac fermions. 
The ellipses in Eq.~(\ref{eq:axionEFTEW})
stand for extra terms like off-diagonal fermion currents, including 
vector ones. 
This Lagrangian is matched with the axion effective Lagrangian below the scale of chiral symmetry breaking,  
$\Lambda_{\chi} \approx 1$ GeV, which reads 
\begin{align}
\label{eq:axionEFTQCD}
\mathcal{L}_a &= -\frac{C_{aN\gamma}}{2m_N} \frac{a}{f_a} \bar N i \gamma_5 \sigma_{\mu\nu} N F^{\mu\nu} 
+ C_{aN} \frac{\partial_\mu a}{2 f_a} \bar N \gamma^\mu \gamma_5 N \nonumber \\
&+ C_{a\pi} \frac{\partial_\mu a}{f_a f_\pi} (2 \partial^\mu \pi^0 \pi^+ \pi^- - \pi^0 \partial^\mu \pi^+ \pi^-  - \pi^0 \pi^+ \partial^\mu \pi^-) \nonumber \\
&+ C_{ae} \frac{\partial_\mu a}{2 f_a} \bar e \gamma^\mu \gamma_5 e 
+ C_{a\gamma} \frac{\alpha}{8\pi} \frac{a}{f_a} F_{\mu\nu} \tilde F^{\mu\nu} + \ldots \, ,
\end{align}
where we kept only terms that are relevant for axion phenomenology, 
namely nucleons ($N = p, n$), pions, electrons and photons. 
In fact, the axion-pion coupling is relevant for the axion hot dark matter bound through to axion-pion thermalization channel
\cite{Chang:1993gm,Hannestad:2005df,DiLuzio:2021vjd} 
while the other couplings are constrained by astrophysical considerations.  
The Wilson coefficients of the two effective Lagrangians in Eqs.~(\ref{eq:axionEFTEW})-(\ref{eq:axionEFTQCD}) 
are related as follows (see e.g.~\cite{DiLuzio:2020wdo})
\begin{align}
\label{eq:CaNgamma}
C_{an\gamma} &= - C_{ap\gamma} = 0.0033(15) \, , \\
\label{eq:Capnsum}
C_{ap} + C_{an} &= (c^0_u + c^0_d -1) (\Delta u + \Delta d) - 2\delta_s \, , \\
\label{eq:Capndiff}
C_{ap} - C_{an} &= (c^0_u - c^0_d - \frac{1 - z}{1 + z}) (\Delta u - \Delta d) \, , \\
\label{eq:Capi}
C_{a\pi} &= -\frac{1}{3} ( c^0_u - c^0_d - \frac{1 - z}{1 + z}) \, , \\
\label{eq:Cae}
C_e &= c^0_e \, , \\
\label{eq:Cagamma}
C_{a\gamma} &= \frac{E}{N} - 1.92(4) \, , 
\end{align}
where $\delta_s = 0.038(5) c^0_s + 0.012(5) c^0_c + 0.009(2) c^0_b + 0.0035 (4)c^0_t$, 
$z \equiv m_u / m_d = 0.48(3)$, $\Delta u + \Delta d = 0.521(53)$ and $\Delta u - \Delta d = 1.2723(23)$ \cite{GrillidiCortona:2015jxo}. 
Here, we neglected corrections coming from flavour mixing as well as radiative corrections (see below). 

The condition that we want to impose corresponds to  
\beq 
\label{eq:stealthcondition}
C_{ap} \approx C_{an} \approx C_{a\pi} \approx C_{ae} \approx C_{a\gamma} \approx 0 \, ,  
\eeq
such that axion phenomenology is driven by the
nucleon EDM couplings. From an effective field theory 
point of view the couplings 
in Eqs.~(\ref{eq:Capnsum})-(\ref{eq:Cagamma}) should be regarded 
as free parameters and hence it is conceivably possible that they are 
suppressed with respect to the $\mathcal{O}(1)$ values suggested 
by benchmark axion models. 

Here, we want to provide a proof of existence of a UV completion that can realize the 
conditions in Eq.~(\ref{eq:stealthcondition}). To this end, we start from the 
non-universal axion model of Ref.~\cite{Bjorkeroth:2019jtx}, which  
can realize the nucleo/pion/electro-phobic conditions 
\beq 
\label{eq:stealthconditionred}
C_{ap} \approx C_{an} \approx C_{a\pi} \approx C_{ae} \approx 0 \, ,   
\eeq
at the price of a \emph{single} tuning. 
The model extends the scalar sector of the SM with three Higgs doublets 
$H_{1,2,3} \sim (1,2,-1/2)$
and a SM singlet $\phi \sim (1,1,0)$. The scalar potential features the non-Hermitian 
SM invariant operators 
\beq 
H_3^\dag H_1 \phi^2 \, , \quad H_3^\dag H_2 \phi^\dag \, , 
\eeq
which imply the conditions (normalizing to the unity the PQ charge of $\phi$, i.e.~$\X_\phi = 1$)
\begin{align}
\label{eq:cond1}
-\X_3 + \X_1 + 2 &= 0 \, , \\
\label{eq:cond2}
-\X_3 + \X_2 - 1 &= 0 \, , \\
\label{eq:cond3}
\X_1 v_1^2 + \X_2 v_2^2 + \X_3 v_3^2 &= 0 \, , 
\end{align}
where the latter condition arises from the orthogonality between the PQ and hypercharge currents, 
with $\langle H_{1,2,3} \rangle = v_{1,2,3}$ and $v^2 = v^2_1 + v^2_4 + v^2_3 \approx (174 \, \text{GeV})^2$ 
the square of the Higgs vacuum expectation value. The Yukawa sector features the following operators, 
with a non-universal assignment of the PQ charges in the quark sector 
with a 2+1 structure (i.e.~first and second family, 
denoted by greek indices,  
are characterized by the same PQ charge)
\begin{align}
\bar q_{\alpha} u_{\beta} H_1 \, , \ \bar q_{3} u_{3} H_2 \, , \ \bar q_{\alpha} u_{3} H_1 \, , \ \bar q_{3} u_{\beta} H_2 \, , \\ 
\bar q_{\alpha} d_{\beta} \tilde H_2 \, , \ \bar q_{3} d_{3} \tilde H_1 \, , \ \bar q_{\alpha} d_{3} \tilde H_2 \, , \ \bar q_{3} d_{\beta} \tilde H_1 \, ,    
\end{align}
with $\tilde H_{1,2} = (i \sigma_2) H^*_{1,2}$.
Differently from Ref.~\cite{Bjorkeroth:2019jtx}, 
that assumed a universal PQ charge assignment 
in the lepton sector, 
in order to obtain here a suppressed coupling to photons 
(see below)
we assume 
\beq 
\bar \ell_{1} e_{1} \tilde H_3 \, , \ \ell_{2} e_{2} \tilde H_1 \, , \ \ell_{3} e_{3} \tilde H_2 \, , \ \ldots \ \, .
\eeq 
Neglecting flavour mixing,\footnote{In the presence of flavour mixing, 
$c^0_f \to c^0_f + \Delta c^0_f$, where $\Delta c^0_f$ involves 
off-diagonal elements of fermion mass diagonalization matrices, 
which are assumed here to be negligible. See Refs.~\cite{DiLuzio:2017ogq,Bjorkeroth:2018ipq} for details.} 
the flavour diagonal axion couplings to the 
axial current read 
\begin{align}
\label{eq:c0u}
c^0_{u,c,t} &= \frac{1}{2N} (\X_{u_{1,2,3}} - \X_{q_{1,2,3}}) \nonumber \\
& =
\{ \frac{2}{3} - \frac{\X_3}{3}, \frac{2}{3} - \frac{\X_3}{3}, - \frac{1}{3} - \frac{\X_3}{3} \}
\, , \\
\label{eq:c0d}
c^0_{d,s,b} &= \frac{1}{2N} (\X_{d_{1,2,3}} - \X_{q_{1,2,3}}) \nonumber \\
& = 
\{ \frac{1}{3} + \frac{\X_3}{3}, \frac{1}{3} + \frac{\X_3}{3}, - \frac{2}{3} + \frac{\X_3}{3} \} \, , \\
\label{eq:c0e}
c^0_{e,\mu,\tau} &= \frac{1}{2N} (\X_{e_{1,2,3}} - \X_{\ell_{1,2,3}}) \nonumber \\
& = 
\{ \frac{\X_3}{3}, - \frac{2}{3} + \frac{\X_3}{3}, \frac{1}{3} + \frac{\X_3}{3} \} \, ,  
\end{align}
where we used the value of the QCD anomaly factor $N$ given by 
\begin{align} 
2 N &= \sum_{i=1}^3 (\X_{u_i} + \X_{d_i} - 2 \X_{q_i}) 
= 3 
\, ,  
\end{align}
and we also used Eqs.~(\ref{eq:cond1})-(\ref{eq:cond2}).
Since, by construction, we have  
\beq 
c^0_{u} + c^0_{d} = 1 \, ,
\eeq
then Eq.~(\ref{eq:Capnsum}) implies $C_{ap} + C_{an} \approx 0$ 
(up to $\mathcal{O}(5\%)$ corrections from $\delta_s$). 
On the other hand, the condition $C_{ap} - C_{an} = 0$ requires 
the tuning
(see Eq.~(\ref{eq:Capndiff}))
\beq 
 c^0_u - c^0_d - \frac{1 - z}{1 + z} = 0 \, ,  
\eeq
that is 
\beq 
\frac{\X_1 + \X_2}{\X_1 - \X_2} 
= \frac{1}{3} - \frac{2}{3} \X_3
= \frac{1 - z}{1 + z} = 0.35 \approx \frac{1}{3} \, , 
\eeq
where in the second step we used 
Eqs.~(\ref{eq:cond1})-(\ref{eq:cond2}). 
Note that this condition 
(satisfied for $\X_3 \approx 0$)
also automatically guarantees $C_{a\pi} = 0$ (see Eq.~(\ref{eq:Capi})) and 
$C_e \approx 0$ (see Eq.~(\ref{eq:Cae}) and Eq.~(\ref{eq:c0e})).
Defining the vacuum angles $\beta_{1,2}$ via 
\begin{align}
v_1 &= v \cos\beta_1 \cos\beta_2 \, , \\
v_2 &= v \sin\beta_1 \cos\beta_2 \, , \\
v_3 &= v \sin\beta_2 \, , 
\end{align}
we can express 
\beq 
\label{eq:X3vacuum}
\X_3 = (3 \cos^2\beta_1 - 1) \cos^2\beta_2 \, , 
\eeq
and parametrize the couplings above in terms of the vacuum angles 
that are subject to perturbative unitarity constraints (see Ref.~\cite{Bjorkeroth:2019jtx}). 

Finally, the QED anomaly factor can be split into a quark plus lepton contribution, 
i.e.~$E = E_Q + E_L$, which read respectively 
\begin{align}
E_Q &= \sum_{i=1}^3 3 \left(\frac{2}{3}\right)^2 (\X_{u_i} - \X_{q_i}) 
+ 3 \left(-\frac{1}{3}\right)^2 (\X_{u_i} - \X_{d_i}) \nonumber \\
&=
4 - 3 \X_3 \, , \\
E_L &= \sum_{i=1}^3 (-1)^2 (\X_{e_i} - \X_{\ell_i}) 
= 3 \X_3 -1 \, ,
\end{align}  
and hence $E / N = 2$, corresponding to the photo-phobic 
coupling $C_{a\gamma} = 0.08(4)$.

In summary, the model is characterized 
by the following axion couplings: 
\begin{align}
\label{eq:CaNgammaS4}
C_{an\gamma} &= - C_{ap\gamma} = 0.0033(15) \, , \\
\label{eq:CapnsumS4}
C_{ap} + C_{an} &= -0.027(3) - 0.021(3) \, \X_3 \, , \\
\label{eq:CapndiffS4}
C_{ap} - C_{an} &= -0.023(35) - 0.848(35) \, \X_3 \, , \\
\label{eq:CapiS4}
C_{a\pi} &= 0.006(9) + 0.222(9) \, \X_3 \, , \\
\label{eq:CaeS4}
C_e &= \frac{\X_3}{3} \, , \\
\label{eq:CagammaS4}
C_{a\gamma} &= 0.08(4) \, . 
\end{align}
The condition in Eq.~(\ref{eq:stealthcondition}) is hence obtained 
at the prize of a \emph{single} tuning, i.e.~$\X_3 \approx 0$, where $\X_3$ can be expressed in terms of vacuum angles of the extended Higgs sector, 
see Eq.~(\ref{eq:X3vacuum}). In particular, $\X_3 = 0$ is obtained for $\cos^2\beta_1 = 1/3$ 
which is fully within the perturbative domain of the model 
(see Ref.~\cite{Bjorkeroth:2019jtx}). 
We remark that the advocated level of suppression of axion couplings can be kept 
also upon including running effects from $f_a$ 
to the QCD scale, albeit within fairly different parameter space regions than in the tree-level case \cite{DiLuzio:2022tyc}

On the other hand, the level of cancellation 
that can be achieved in the model above 
(for $\X_3 \approx 0$)
is not yet sufficient to make the nucleon-EDM axion 
coupling the most important one for axion phenomenology. Indeed, the two strongest astrophysical constraints on $f_a$ are due to $C_{a\gamma}$ and $C_{aN}$, coming from HB stars~\cite{Ayala:2014pea} and SNe~\cite{Carenza:2019pxu}, respectively. An order of magnitude suppression in $C_{a\gamma}$ is sufficient to evade the former, while the latter is a factor $\sim 800$ stronger than the SN bound due to nucleon-EDM coupling. Therefore the suppression proposed in the model in Eqs.~\eqref{eq:CaNgammaS4}-\eqref{eq:CagammaS4} is enough to evade the HB bound but not sufficient for the SN bound due to $C_{aN}$. Hence a further order of magnitude cancellation 
in $C_{aN}$
would be required. 
This can be achieved at the price of extra tunings.  
For instance, $C_{aN}$ can be further suppressed 
by taking into account flavour mixing effects for 
flavour-diagonal axion couplings (see Refs.~\cite{DiLuzio:2017ogq,Bjorkeroth:2018ipq} for details), while $C_{a\gamma}$ can be modified via an extra 
KSVZ-like fermionic 
sector along the lines of \cite{DiLuzio:2016sbl,DiLuzio:2017pfr}
which contributes to the electromagnetic anomaly. 

We conclude that although a percent level suppression of axion couplings seems perfectly feasible in explicit QCD axion models, 
going below that level requires further non-trivial assumptions.

\section{Evaluation of the emissivity and absorption mean free path}
\label{sec:appem}

The axion emissivity due to Compton scattering $N+\gamma \rightarrow N + a$ is given by
\begin{equation}
\begin{split}
Q_{a}&= \sum_{\rm nucleons} \int \frac{2 d^3 \bp_i}{(2\pi)^3 2 E_i}\frac{2 d^3 \bp_f}{(2\pi)^3 2 E_f}\frac{3 d^3 \bk}{(2\pi)^3 2 E_k} \frac{ d^3 \bp_a}{(2\pi)^3 2 E_a}\\
    &E_{a}(2\pi)^4\delta^{4}(P_{i}+K-P_{f}-P_{a})|\overline{\mathcal{M}}|^2 f_{p_i} f_{k} (1-f_{p_f})\,,
    \end{split}
\label{eq:qapp}
\end{equation}
where $|\overline{\mathcal{M}}|^2$ is given by Eq.~\eqref{eq:sqmSN}, $P_i=(E_i,\,\textbf{p}_i)$, $K=(E_k,\,\textbf{k})$, $P_f=(E_f,\,\textbf{p}_f)$ and  $P_a=(E_a,\,\textbf{p}_a)$ are the 4-momenta of the initial and final state nucleon $N$, the photon and the axion, respectively. In addition, $f_i$'s are the usual Fermi-Dirac or Bose-Einstein distributions, i.e.
\begin{align}
f_i(E)=\frac{1}{e^{\left[ E_i(p_i)-\mu_i\right] /T} \pm1}\;,
\end{align}
where the $+$ sign applies to fermions, the $-$  is for bosons, and $\mu_i$ are the chemical potentials for $i=p,n$, while photons have vanishing chemical potential. Following Ref.~\cite{Hannestad:1995rs}, the integration in Eq.~\eqref{eq:qapp} can be simplified and the emissivity is given by
\begin{equation}
\begin{split}
    Q_a (E_a) &= \sum_{\rm nucleons} \frac{3 g_{d}^2}{2^5 \pi^6} \int_{m_a}^\infty dE_a \int_0^{\infty} dp_f \int_{0}^{p_a+p_f} dk \\ 
    & \int_{\alpha}^{\beta} d\cos\theta   \frac{p_f^2}{E_f} \frac{k^2}{E_k}  p_a E_a (I_0 + I_1 + I_2 ) f_k f_{p_i} (1- f_{p_f})\,,
\end{split}
\end{equation}
where $p_a=|\textbf{p}_a|$, $k=|\textbf{k}|$, $p_f=|\textbf{p}_f|$ and 
\begin{equation}
\begin{split}
I_0=& \bigg[\frac{4}{3}(E_a\,E_f-E_a\,E_k+p_a\,k\cos\theta + Q/2)\times \\
& \times(E_a\,E_f-m_\gamma^2+Q/2)\,+m_N^2\,m_\gamma^2- \\
& -\frac{m_\gamma^2}{3}\,(E_a\,E_k-p_a\,k\,\cos\theta+m_N^2-Q/2) \bigg] \frac{\pi}{\sqrt{-a}}\,,\\
I_1&=\frac{4}{3}\big[p_a\,p_f\,(E_a\,E_f-m_\gamma^2+Q/2)+\\
&+p_a\,p_f\,(E_a\,E_f-E_a\,k+p_a\,k\,\cos\theta+Q/2)\big]\,\frac{b}{2\,a}\,\frac{\pi}{\sqrt{-a}}\,,\\
I_2&=\frac{4}{3}p_a^2\,p_f^2\,\left(\frac{3\,b^2}{8\,a^2}-\frac{c}{2\,a}\right)\frac{\pi}{\sqrt{-a}}\,,\\
\end{split}
\end{equation}
with $\theta$ the angle between the axion and the photon momenta, and $Q$, $a$, $b$, $c$ given by
\begin{equation}
\begin{split}
Q=& m_a^2+m_N^2+m_\gamma^2-m_N^2\,,\\
a=& p_f^{2}\,(-4\,\kappa+8\,\epsilon)\,,\\
b=& p_f\,(p_a-\epsilon/p_a)\,(8\,\gamma+4\,Q+o\,\epsilon)\,,\\
c=& -4\,\gamma^2-4\,\gamma\,Q-Q^2-8\,\gamma\,\epsilon-\\
&- 4\,Q\,\epsilon-4\epsilon^2 + 4\,p_i^2\,k^2\,(1-\cos\theta^2)\,,\\
\end{split}
\end{equation}
where the constants $\kappa$, $\gamma$, $\epsilon$ are
\begin{equation}
\begin{split}
 \kappa =& p_a^2 + k^2\,,\\
 \gamma =& E_a\,E_f - E_a\,E_k - E_f\,E_k\,,\\
 \epsilon =& p_a\,k\,\cos\theta\,.
 \end{split}
\end{equation}
The integration domain for $\cos\theta$ is composed by $\alpha=\sup[-1,\cos\theta_{\min}]$ and $\beta=\inf[+1,\cos\theta_{\max}]$, with $\alpha\lesssim \beta$ and 
\begin{equation}
\begin{split}
 \cos\theta_{\max\,,\min} & =  \frac{-2\gamma -2 p_f^2 - Q }{2\,p_a\,k}\pm \\
 & \pm \frac{2 \,p_f \sqrt{2\gamma +p_a^2 + p_f^2 + k^2 + Q}}{2\,p_a\,k}\,.
 \end{split}
\end{equation}
Finally, we stress that in the case of interest $p_a\sim E_a$, since we are considering light axions ($m_a\ll E_a$). 

On the other hand, axions may be absorbed due to the inverse process $a + N \rightarrow \gamma + N$. The absorption mean free path $\lambda$ is
\begin{equation}
    \lambda^{-1} (E_a)=n_N\,\sigma_{aN\rightarrow \gamma N} (E_a)
    \label{eq:lambdas}\,
\end{equation}
where $n_N$ is the nucleon density and $\sigma_{aN\rightarrow \gamma N}(E_a)$ is the absorption cross section given by
\begin{equation}
\begin{split}
\sigma_{aN\rightarrow \gamma\,N}(E_{a})&=\frac{1}{n_{N}}\frac{1}{2\,E_a}\\
&\int \frac{2 d^3p_i}{2E_i(2\pi)^3}\frac{2 d^3p_f}{2E_f(2\pi)^3}\frac{3 d^3k}{2\omega(2\pi)^3}\\
    &(2\pi)^4 \delta^{4}(P_a+P_f-K-P_i) |\overline{\mathcal{M}}|^2 \\ &f_{p_f}(1-f_{p_i})(1+f_k)\,\, .
    \end{split}
\label{eq:crossect}
\end{equation}
This expression differs from Eq.~\eqref{eq:qapp} due to the absence of the integration over the axion energy, i.e. $d^3\textbf{p}_a/(2\pi)^3\,E_a$, and the interchange between the initial and final states. Therefore, following once more Ref.~\cite{Hannestad:1995rs}, one obtains
\begin{equation}
\begin{split}
    \lambda^{-1} (E_a) &= \frac{3 g_{d}^2}{(2\pi)^4\,E_a} \int_{0}^{\infty} dp_f \int_{0}^{p_a+p_f} dk \int_{\alpha}^{\beta} d\cos\theta \\
    & \frac{p_f^2}{E_f} \frac{k^2}{E_k} (I_0 + I_1 + I_2 ) f_{p_f} (1+f_k) (1-f_{p_i})\,.
\end{split}
 \label{eq:lambda}
\end{equation}

\section{Emissivities in the non-degenerate and non-relativistic limit}
\label{sec:appapprox}

\subsection{Compton effect}
In order to obtain a simple expression for the Compton emissivity, let us evaluate it in the non-degenerate and non-relativistic limit for nucleons, ignoring also the effective photon mass. The general expression for the emissivity is
\begin{equation}
\begin{split}
Q_{a}&= \sum_{\rm nucleons} \int \frac{2 d^3 \bp_i}{(2\pi)^3 2 E_i}\frac{2 d^3 \bp_f}{(2\pi)^3 2 E_f}\frac{2 d^3 \bk}{(2\pi)^3 2 E_k} \frac{ d^3 \bp_a}{(2\pi)^3 2 E_a}\\
    &E_{a}(2\pi)^4\delta^{4}(P_{i}+K-P_{f}-P_{a})|\overline{\mathcal{M}}|^2 f_{p_i} f_{k} (1-f_{p_f})\,,
    \end{split}
\label{eq:qappapprox}
\end{equation}
where $|\overline{\mathcal{M}}|^2=2\,g_{d}^2\,(K\cdot P_i)\,(K\cdot P_f)$, $P_i\approx(E_i,\,\textbf{p}_i)$, $P_f\approx(E_f,\,\textbf{p}_f)$, $K=(E_k,\,\textbf{k})$ and  $P_a=(E_a,\,\textbf{p}_a)$ are the 4-momenta of the initial and final state nucleon $N$, the photon and the axion, respectively, with $E_{i,f}\approx |\bp_i|^2/2m_N+m_N$, $E_k = |\bk| \equiv k$ and $E_a = |\bp_a| \equiv p_a$. In addition, all the $f_i$'s are considered as Maxwell-Boltzmann distributions, i.e.
\begin{align}
f_i(E)=e^{-\left( E_i-\mu_i\right) /T}\;,
\end{align}
where $\mu_i$ are the chemical potentials for nucleons, while photons and axions have vanishing chemical potential. In the non degenerate limit the Pauli blocking factors are negligible, $(1-f_{p_f})\approx 1$, and we can approximate
\begin{equation}
    f_{p_i}\,f_{k} \approx  f_{p_f}\,f_{p_a} = e^{-E_a/T}\,e^{-(E_f-\mu)/T}\,.
\end{equation}
To further simplify the expression, let us fix as reference system the center of momentum frame, in which $\bp_i = - \bk$ and $\bp_f = - \bp_a$. We can integrate over $p_i$ to eliminate $\delta^{4}(P_{i}+K-P_{f}-P_{a})$, obtaining a constraint on $k$, since
\begin{equation}
    \delta(P_i^2 - m_N^2) = \frac{1}{2(E_f + E_a)}\,\delta (k-E_a)
\end{equation}
and rewrite the transition matrix as
\begin{equation}
|\overline{\mathcal{M}}|^2= 2\,g_{d}^2\,k^2\,(E_f+E_a)\,(E_f+E_a\,z)\,,
\end{equation}
being $\bp_a \cdot \bk = |\bp_a|\, |\bk|\,z$. Since $d^3\bk = 4\pi\,dk\,k^2$, $d^3\bp_a = 2\pi\,dE_a\,E_a^2\,dz$ and $d^3\bp_a = 4\pi\,dp_f\,p_f^2$, we obtain
\begin{eqnarray}
    Q_a =&& \frac{g_{d}^2}{2^3\,\pi^5}\int_{-1}^{+1}dz\,\int\,dE_a E_a^5 \int dp_f\,\frac{p_f^2}{E_f}\times   \nonumber \\
    && e^{-E_a/T}\,e^{-(E_f-\mu)/T}\,(E_f+E_a\,z)\,.
\end{eqnarray}
Integrating over $z$, and exploiting the relation in the non-degenerate and non-relativistic limit
\begin{equation}
    e^{(E_f-\mu)/T}=\frac{\rho Y_N}{2 m_N} \left(\frac{2\pi}{m_N T}\right)^{1.5}\,e^{-p_f^2/2\,m_N\,T}
    \label{eq:munondeg}
\end{equation}
we obtain
\begin{eqnarray}
    Q_a &=& \frac{g_d^2}{2^3\,\pi^5}\,\frac{\rho\,Y_N}{m}\,\left(\frac{2\pi}{m_N T}\right)^{1.5}\,\times  \nonumber \\ 
    &&\int dE_a\,E_a^5\,e^{-E_a/T}\,\int dp_f\,p_f^2\,e^{-p_f^2/2\,m_N\,T}\,.
\end{eqnarray}
Now, using the relations $\int dE_a\,E_a^5\,e^{-E_a/T}=120\,T^2$ and $\int dp_f\,p_f^2\,e^{-p_f^2/2\,m_N\,T}=\sqrt{\pi/2}\,(m_N\,T)^{1.5}$, we conclude that
\begin{equation}
    Q_a = \rho\,Y_N\,\frac{30\,g_d^2}{\pi^3}\,\frac{T^6}{m_N}\,.
\end{equation}
We are considering both the contributions of protons and neutron $Y_N = Y_e + Y_n \approx 1$, then the emissivity per unit mass $\varepsilon_a = Q_a/\rho$ can be written in the following form
\begin{eqnarray}
    \varepsilon_a &=& \frac{30\,g_d^2}{\pi^3}\,\frac{T^6}{m_N} \approx 10^{36} \erg\, \g^{-1} \s^{-1} \times \nonumber \\
    &&\left(\frac{g_d}{\GeV^{-2}}\right)^2\,\left(\frac{T}{30\,\MeV}\right)^6\,\left(\frac{938\,\MeV}{m_N}\right)\,. \nonumber \\
\label{eq:qacompt}
\end{eqnarray}
For typical SN conditions ($\rho=3\times 10^{14}$ g cm$^{-3}$, $T=30$~MeV, $Y_e=0.3$), imposing the cooling bound $\varepsilon < 10^{19}$~erg g$^{-1}$ s$^{-1}$, the coupling is constrained to be $g_{d}\lesssim 3 \times 10^{-9}$~GeV$^{-2}$, in rough agreement with the estimation in Ref.~\cite{Graham:2013gfa}.

\subsection{Bremsstrahlung}

In this Section, we evaluate the bremsstrahlung emissivity in the non-relativistic and non-degenerate approximation for nucleons, which is thought to be a good approximation in a SN core, and we show that this process is subdominant with respect to the Compton emission for typical SN conditions.

\begin{figure}[t!]
\vspace{0.cm}
\includegraphics[scale=0.78]{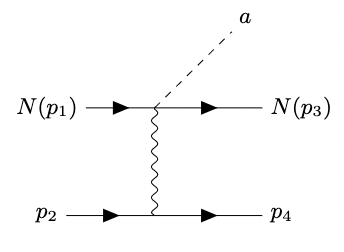}
\caption{Feynman diagram for the $np$ bremsstrahlung.}
\label{fig:feynmnp}
\end{figure}

\begin{figure}[t!]
\vspace{0.cm}
\includegraphics[scale=0.78]{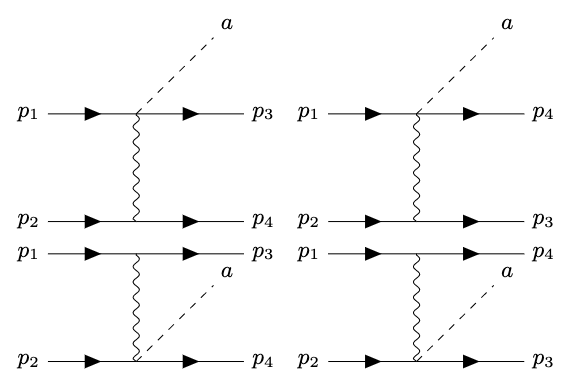}
\caption{Feynman diagrams for the $pp$ bremsstrahlung process. The left panels show the direct diagrams (type $a$), while the right ones show the exchange diagrams (type $b$).}
\label{fig:feynmpp}
\end{figure}

In the bremsstrahlung process, an axion is produced after the interaction between a nucleon ($N_1=N_3$) and a virtual photon emitted by a proton ($N_2=N_4=p$). We classify the bremsstrahlung as $np$ channel if $N_1=N_3=n$ (see Fig.~\ref{fig:feynmnp}) and $pp$ channel if $N_1=N_3=p$ (see Fig.~\ref{fig:feynmpp}). Therefore the emissivity is given by the sum of the two channels
\begin{equation}
    Q_{a,B} = Q_{a,np} + Q_{a,pp}\,,
    \label{eq:Qabremtot}
\end{equation}
where $Q_{a,np}$ is due to $np$ process and $Q_{a,pp}$ is due to the $pp$ channel. Let us start from the $np$ process. In this case the emissivity is given by

\begin{eqnarray}
Q_{a,np} & = & \int  \frac{2 d^3 \bp_1}{(2\pi)^3 2 E_1} \frac{2 d^3 \bp_2}{(2\pi)^3 2 E_2} \frac{2 d^3 \bp_3}{(2\pi)^3 2 E_3}\frac{2 d^3 \bp_4}{(2\pi)^3 2 E_4}\nonumber \\
  && \frac{ d^3 \bp_a}{(2\pi)^3 2 E_a} E_a\,(2\pi)^4\delta^{4}(p_{1}+p_{2}-p_{3}-p_{4}-p_a) \nonumber \\
&& \times |\overline{\mathcal{M}}_{np}|^2 f_{1} f_{2} (1-f_{3})(1-f_{4})\,, 
 \label{eq:emissnp}
\end{eqnarray}
with
\begin{equation}
 |\overline{\mathcal{M}}_{np}|^2 = \frac{1}{16} \sum_{\rm{spins}} |\mathcal{M}_{np}|^2  \,,
\end{equation}
where we are averaging the nucleon spins of the initial and final states and the matrix element is (see Fig.~\ref{fig:feynmnp})
\begin{widetext}
\begin{equation}
  \mathcal{M}_{np}=\frac{e\,g_{d}}{4} \bar{u}(p_4) \gamma^\alpha\, u(p_2)\,\frac{k^\mu g_{\alpha \nu}-k^\nu\,g_{\alpha \mu}}{k^2}\,\bar{u}(p_3)\,(\gamma^\mu\gamma^\nu - \gamma^\nu \gamma^\mu)\gamma^5\,u(p_1)\,,
 \label{eq:M}
\end{equation}
\end{widetext}
being $e$ the electric charge, $k=p_2-p_4$ the transferred-photon 4-momentum and $g^{\mu\nu}$ the metric tensor. Assuming that nucleons are non-relativistic and non-degenerate, $(1-f_3)(1-f_4)\approx 1$, we can write
\begin{eqnarray}
Q_{a,\,np} &=& \int \frac{2 d^3 \bp_1}{(2\pi)^3 2 m_N} \frac{2 d^3 \bp_2}{(2\pi)^3 2 m_N} \frac{2 d^3 \bp_3}{(2\pi)^3 2 m_N}\frac{2 d^3 \bp_4}{(2\pi)^3 2 m_N} \nonumber \\
  &&\frac{ d^3 \bp_a}{(2\pi)^3 2 E_a}  E_a\,(2\pi)^4\delta^{4}(p_{1}+p_{2}-p_{3}-p_{4}-p_a) \nonumber \\
  && |\overline{\mathcal{M}}_{np}|^2 f_{1} f_{2}\,, 
 \label{eq:emissND}
\end{eqnarray}
where
\begin{equation}
    f_i = \frac{n_i}{2}\left(\frac{2\pi}{m_N\,T}\right)^{3/2}\,e^{-\textbf{p}_i^2/(2\,m_N\,T)}\,,
\end{equation}
with $i=p,n$, and $n_p = \rho Y_e/ m_N$ and $n_n = \rho (1-Y_e)/ m_N$. In addition, due to the non-relativistic approximation
\begin{eqnarray}
p_i\approx \left(m_N+\frac{\textbf{p}_i^2}{2\,m_N}\,,\textbf{p}_i\right)\,,\\ \nonumber
p_a=(E_a,\textbf{p}_a)\,,\,\quad |\textbf{p}_a|=E_a\,.
\end{eqnarray}
Following Refs.~\cite{Brinkmann:1988vi,Raffelt:1993ix,Giannotti:2005tn,Dent:2012mx} let us introduce the center-of-momentum variables
\begin{eqnarray}
\textbf{p}_1&=&\textbf{P}+\textbf{p}_i\,\quad\, \textbf{p}_2=\textbf{P}-\textbf{p}_i \nonumber \\
\textbf{p}_3&=&\textbf{P}'+\textbf{p}_f\,\quad\, \textbf{p}_4=\textbf{P}'-\textbf{p}_f\,. 
 \label{eq:PP'}
\end{eqnarray}
In the non-relativistic limit, a typical nucleon with kinetic energy $E_{\rm kin}$ has momentum $|\textbf{p}_i| = \sqrt{2\,m_N\,E_{\rm kin}}\gg E_{\rm kin} \approx E_a$. Then, the three dimensional delta function implies $\textbf{P}=\textbf{P}'$. Due to the non relativistic approximation
\begin{equation}
    p_i\cdot p_j = m_N^2 + \frac{1}{2} (\textbf{p}_i-\textbf{p}_j)^2\,,
\end{equation}
and using Eq.~\eqref{eq:PP'}\,, we obtain
\begin{eqnarray}
p_1\cdot p_2 &=& m_N^2 + 2 |\textbf{p}_i|^2\,, \nonumber \\
p_3\cdot p_4 &=& m_N^2 + 2 |\textbf{p}_f|^2\,, \nonumber\\
p_1\cdot p_3 &=& m_N^2 + \frac{1}{2}\left[|\textbf{p}_i|^2+|\textbf{p}_f|^2-2\textbf{p}_i\cdot \textbf{p}_f\right]\,, \nonumber\\
p_1\cdot p_4 &=& m_N^2 + \frac{1}{2}\left[|\textbf{p}_i|^2+|\textbf{p}_f|^2+2\textbf{p}_i\cdot \textbf{p}_f\right]\,, \\
p_2\cdot p_3 &=& m_N^2 + \frac{1}{2}\left[|\textbf{p}_i|^2+|\textbf{p}_f|^2+2\textbf{p}_i\cdot \textbf{p}_f\right]\,, \nonumber\\
p_2\cdot p_4 &=& m_N^2 + \frac{1}{2}\left[|\textbf{p}_i|^2+|\textbf{p}_f|^2-2\textbf{p}_i\cdot \textbf{p}_f\right]\,. \nonumber
\end{eqnarray}
Then, given $|\textbf{P}|=P$, $|\textbf{p}_i|=p_i$, $|\textbf{p}_f|=p_f$, $\textbf{p}_i\cdot \textbf{p}_f = p_i p_f z$, the squared matrix element can be written in terms of the new variables as 
\begin{widetext}
\begin{eqnarray}
    |\overline{\mathcal{M}}_{np}|^2&=&\frac{g_{d}^2\,4\pi\alpha}{(p_f^2-2\,p_f\,p_i\,z+p_i^2)^2}\bigg\{m_N^2\bigg[6\,p_f^2\,p_i^2(3-2z^2)-4\,p_f^3\,p_i\,z+p_f^4-4\,p_f\,p_i^3\,z+p_i^4\bigg]\\ \nonumber &+&4m_N^4(p_f^2-2\,p_f\,p_i\,z+p_i^2)-8\,p_f^2\,p_i^2\,z^2(p_f^2+p_i^2)+2(p_f^2+p_i^2)^3\bigg\}
\end{eqnarray}
\end{widetext}
and 
\begin{equation}
    f_1\,f_2 = \frac{n_n\,n_p}{4}\left(\frac{2\pi}{m_N\,T}\right)^3\,e^{-p_i^2/(m_N\,T)}\,e^{-P^2/(m_N\,T)}\,.
\end{equation}
One can introduce the variables
\begin{equation}
    u=\frac{|\bp_i|^2}{m_N\,T}\,,\quad\,u=\frac{|\bp_f|^2}{m_N\,T}\,,\quad\,x=\frac{E_a}{T}\,,
\end{equation}
then the delta function in Eq.~\eqref{eq:emissND} can be rewritten as
\begin{widetext}
\begin{equation}
  \delta^{(4)}(p_1+p_2-p_3-p_4-p_a)=  \delta^{(3)}(\bp_1+\bp_2-\bp_3-\bp_4-\bp_a)\frac{\delta(u-v-x)}{T}
\end{equation}
\end{widetext}
and we integrate over $d^3\bp_4$ using the three-dimensional piece. We now change the integration variables $d^3 \bp_1\,d^3 \bp_2\,d^3 \bp_3=8\,d^3 \bP\,d^3 \bp_i\,d^3 \bp_f$\, where $8$ is due to the Jacobian, and using
\begin{eqnarray}
    \int d^3 \bP e^{-P^2/(m_N\,T)} &=&  (\pi\,m_N\,T)^{3/2}\,,  \\
    \int d^3\,\bp_i e^{-p_i^2/m_N T} &=& 2\pi (m_N\,T)^{3/2}\int du e^{-u}\sqrt{u}\,, \nonumber \\
    \int d^3\,\bp_f e^{-p_i^2/m_N T} &=& \pi (m_N\,T)^{3/2}\int dv \sqrt{v}\,\int_{-1}^{1} dz\,, \nonumber \\
    \int d^3\,\bp_a &=& 4\pi T^3\int dx x^2\,, \nonumber 
\end{eqnarray}
we obtain 
\begin{widetext}
\begin{equation}
 Q_{a,{np}} = \frac{\rho^2\,(1-Y_e)\,Y_e}{32\,\pi^{7/2}}\frac{T^{7/2}}{m_N^{9/2}} \int_0^{\infty} dv \int_0^{\infty} dx \int_{-1}^{+1} dz\,e^{-(v+x)}\sqrt{v+x}\sqrt{v}\,x^2\,|\overline{\mathcal{M}}_{np}|^2\big|_{v,x,u=v+x}\,,  
 \label{eq:Qanp}
\end{equation}
\end{widetext}
where we fix $u=v+x$ due to the $\delta$-function and 
\begin{widetext}
\begin{eqnarray}
   && |\overline{\mathcal{M}}_{np}|^2\big|_{v,x,u=v+x}=\frac{g_{d}^2\,4\pi\alpha}{(m_N\,T\,(2v+x)-2\,z\sqrt{m_N\,T\,v}\,\sqrt{m_N\,T(v+x)})^2}\times \nonumber \\
    && \bigg\{m_N^3\bigg[4m_N^2\,T\,(2v+x)+m_N\,T^2(20v^2-12\,v\,z^2\,(v+x)\,+20\,v\,x+x^2)-2\,T\,(2v+x)\times \nonumber \\ &&[2z\sqrt{m_N\,T\,v}\sqrt{m_N\,T(v+x)}+T^2(4\,v\,z^2(v+x)-(2v+x)^2)]-8\,m_N\,z\sqrt{m_N\,T\,v}\sqrt{m_N\,T(v+x)}\big]\bigg\}\,.
\end{eqnarray}
\end{widetext}
The $pp$-channel contribution (see Fig.~\ref{fig:feynmpp}) is given by
\begin{eqnarray}
Q_{a,pp} = && S\,\int \frac{2 d^3 \bp_1}{(2\pi)^3 2 E_1} \frac{2 d^3 \bp_2}{(2\pi)^3 2 E_2} \frac{2 d^3 \bp_3}{(2\pi)^3 2 E_3}\frac{2 d^3 \bp_4}{(2\pi)^3 2 E_4} \nonumber \\
&&\frac{d^3 \bp_a}{(2\pi)^3 2 E_a}
 E_a\,(2\pi)^4\delta^{4}(p_{1}+p_{2}-p_{3}-p_{4}-p_a) \nonumber \\
 && \times |\overline{\mathcal{M}}_{pp}|^2 f_{1} f_{2} (1-f_{3})(1-f_{4})\,,  
 \label{eq:emisspp}
\end{eqnarray}
where
\begin{equation}
|\overline{\mathcal{M}}_{pp}|^2=\frac{1}{16}|\mathcal{M}_{pp}|^2\,,
\end{equation}
being
\begin{equation}
|\mathcal{M}_{pp}|^2=|\mathcal{M}_a|^2+|\mathcal{M}_b|^2-(\mathcal{M}_a\,\mathcal{M}_b^*+\mathcal{M}_b\,\mathcal{M}_a^*)\,,
\label{eq:Mpp}
\end{equation}
with $a$ the direct-diagram contribution (upper left panel in Fig.~\ref{fig:feynmpp}) and $b$ the exchange-diagram, obtained interchanging the final fermion lines (upper right panel in Fig.~\ref{fig:feynmpp}). In Eq.~\eqref{eq:qapp}, $S$ is the symmetry factor
\begin{equation}
    S=2\times \frac{1}{4}=\frac{1}{2}\,,
\end{equation}
where $2$ comes from the position where the axion can be attached (upper or lower vertex, see the lower panels of Fig.~\ref{fig:feynmpp}), and $1/4$ comes from the identical particles in the initial and final states (for the $np$ process $S=1$). Assuming non-relativistic and non-degenerate nucleons, the matrix element in Eq.~\eqref{eq:Mpp} can be evaluated following a procedure analogous to the $np$ process and the $pp$-contribution reads as
\begin{widetext}
\begin{equation}
 Q_{a,{pp}} = \frac{\rho^2\,Y_e^2}{64\,\pi^{7/2}}\frac{T^{7/2}}{m_N^{9/2}} \int_0^{\infty} dv \int_0^{\infty} dx \int_{-1}^{+1} dz\,e^{-(v+x)}\sqrt{v+x}\sqrt{v}\,x^2\,|\overline{\mathcal{M}}_{pp}|^2\big|_{v,x,u=v+x}  \,,
 \label{eq:Qappfin}
\end{equation}
\end{widetext}
which differs from Eq.~\eqref{eq:Qanp} due to the replacements $(1-Y_e)Y_e \rightarrow Y_e^2$ (since only protons are involved), $32 \rightarrow 64$ in the denominator (due to the symmetry factor $S=1/2$) and $|\overline{\mathcal{M}}_{np}|^2\big|_{v,x,u=v+x}\rightarrow |\overline{\mathcal{M}}_{pp}|^2\big|_{v,x,u=v+x}$, where
\begin{widetext}
\begin{eqnarray}
   && |\overline{\mathcal{M}}_{pp}|^2\big|_{v,x,u=v+x}=\frac{g_{d}^2\,4\pi\alpha}{2\,T ((2v+x)^2-4\,v\,z^2(v+x))^2}\times \nonumber \\
    && \bigg\{m_N\bigg[(2v+x)^2\,(4\,m_N^2\,(4\,v+x)+7\,m_N\,T\,(4v^2+4\,v\,x-x^2)+4\,T^2\,(8\,v^2\,x+4\,v^3+7\,v\,x^2+x^3))+ \nonumber \\
    && 16\,v\,z^2\,(v+x)\,(m_N^2(-(4v+x))+2\,m_N\,T\,v\,(v+x))-16\,T\,v^2\,z^4\,(v+x)^2\,(9m_N+4\,T(3v+x)\bigg]\bigg\}\,.
\end{eqnarray}
\end{widetext}
The total emissivity in Eq.~\eqref{eq:Qabremtot} is obtained summing Eqs.~\eqref{eq:Qanp} and \eqref{eq:Qappfin}. Assuming typical SN conditions ($\rho=3\times 10^{14}$ g cm$^{-3}$, $T=30$~MeV, $Y_e=0.3$), the emissivity per unit mass $\varepsilon_{a,B}=(Q_{a,np}+Q_{a,pp})/\rho \approx 4.5\times 10^{34}$~erg~g$^{-1}$~s$^{-1}$~$(g_{d}/\GeV^{-2})^2$, more than one order of magnitude smaller than the Compton emissivity [see Eq.~\eqref{eq:qacompt}], in agreement with other cases discussed in literature (see e.g. Sec. II C in Ref.~\cite{Caputo:2021rux}).

\bibliographystyle{bibi}
\bibliography{biblio.bib}

\end{document}